\documentclass[a4paper,12pt]{article}
\usepackage{graphicx,amsmath,amsfonts,amssymb,cite}
\usepackage[dvips]{color}
\newcommand{\mj}{m_j}
\newcommand{\az}{\alpha Z}
\newcommand{\me}{m_e}
\newcommand{\mpr}{m_p}
\newcommand{\mf}{M_F}
\newcommand{\mi}{M_I}
\newcommand{\ga}{\gamma}
\newcommand{\ka}{\kappa}
\newcommand{\eps}{\epsilon}
\newcommand{\mub}{\mu_0}
\newcommand{\mun}{\mu_N}
\newcommand{\la}{\langle}
\newcommand{\ra}{\rangle}
\newcommand{\hfs}{hyperfine structure}

\newcommand{\vecb}{\vec{B}}

\newcommand{\De}{\Delta}
\newcommand{\Ga}{\Gamma}
\newcommand{\veps}{\varepsilon}
\newcommand{\dhfs}{\Delta E_{\rm HFS}}
\newcommand{\git}{g_I^{\prime}}
\newcommand{\Ft}{F^{\prime}}
\newcommand{\tv}{\tilde{v}}
\newcommand{\tc}{\tilde{c}}
\newcommand{\om}{\omega}
\newcommand{\bv}{\breve{v}}
\newcommand{\bc}{\breve{c}}
\begin{document}
\begin{center}
\textbf{\large
{ Zeeman effect of the hyperfine structure 
levels in lithiumlike ions}}
\\ \vskip 1cm{} D.~L.~Moskovkin$^1$, 
V.~M.~Shabaev$^1$, and W.~Quint$^2$
\\ \vskip 0.1cm{}
\emph{$^1$ Department of Physics, St. Petersburg State
University, Oulianovskaya 1, Petrodvorets, St.~Petersburg
198504, Russia 
\\ $^2$Gesellschaft f\"{u}r Schwerionenforschung,
Planckstrasse 1, D-64291 Darmstadt, Germany}
\end{center}
\vskip 0.5cm{}

\begin{abstract}
The fully relativistic theory of the Zeeman splitting of the $(1s)^2 2s$
hyperfine-structure levels in lithiumlike ions with $Z=6 - 32$ 
is considered for
the magnetic field magnitude in the range from 1 to 10 T.
The second-order corrections to the Breit -- Rabi 
formula are calculated and discussed 
including the one-electron contributions as well 
as the interelectronic-interaction effects of order $1/Z$.
The $1/Z$ corrections are evaluated within a rigorous QED approach.
These corrections are combined with other interelectronic-interaction, 
QED, nuclear recoil, and nuclear size corrections to obtain high-precision 
theoretical values for the Zeeman splitting in Li-like ions with 
nonzero nuclear spin. The results can be used for a precise
determination of nuclear magnetic moments from $g$-factor experiments.
\begin{flushright}
PACS number(s): 32.60.+i, 31.30.Gs, 31.30.Jv, 12.20.Ds
\end{flushright}
\end{abstract}

\section{Introduction}

Recent measurements of the $g$ factor of hydrogenlike carbon and
oxygen have reached an accuracy of about $2\cdot 10^{-9}$
\cite{her00,hae00,ver04}. The experiments were performed on a single 
hydrogenlike ion confined in a Penning trap with a strong magnetic 
field ($B=3.8$ T).
These measurements considerably stimulated theoretical investigations 
of this effect \cite{blu97,per97,bei00,cza01,kar01a,kar01b,sha01,
mar01,gla02,bei02,sha02a,yer02a,sha02b,nef02,kar02,
mosk04,mosk06,pach05,jent06}. Besides a new
possibility for tests of the  magnetic sector of quantum
electrodynamics (QED), these investigations have already provided
a new determination of the electron mass (see Refs. \cite{ver04,mota05} 
and references therein). Extensions of these experiments to systems
with higher nuclear charge number $Z$ and to ions with nonzero
nuclear spin would also provide the basis for new determinations
of the fine-structure constant \cite{kar01a,wer01,shagla06}, 
the nuclear magnetic moments \cite{wer01}, and the nuclear charge radii. 

Extending theoretical description from an H-like to a Li-like
ion, one encounters a serious complication due to the presence of
additional electrons. A number of relativistic calculations of the 
$g$ factor of Li-like ions were carried out previously 
\cite{heg75,ves80,lin93, yan01, ind01}.
However, to reach the accuracy comparable to the one for
H-like ions, a systematic quantum electrodynamic (QED) treatment 
is required \cite{sha02b,shagla03,glasha04,glavol06,shaand06,mosk07os}. 

For both H- and Li-like heavy ions with nonzero nuclear spin 
the ground-state Zeeman splitting caused by the magnetic field 
in the range from 1 to 10 T is much smaller than the hyperfine
splitting and, therefore, the consideration can be
conveniently reduced to the g factor value \cite{mosk04,mosk07os}. 
However, for H-like ions with $Z=1-20$, 
which are being under current experimental 
investigations at Mainz University, the Zeeman splitting is 
comparable with the hyperfine splitting if the magnitude of 
the homogeneous magnetic field does not exceed 10~T. 
This requires constructing the perturbation
theory for degenerate states. To a good accuracy, the
solution of the problem is given by the well-known Breit -- Rabi
formula \cite{BR31, HKopf, bet57, zap79}. The aforesaid experimental
precision  has, however, shown the need for an improvement
of the Breit -- Rabi formula for H-like ions \cite{mosk06}.

In the present paper, we consider the Breit -- Rabi
formula for the $2s$ hyperfine-structure levels in lithiumlike ions. 
Evaluations of the coefficients of this formula 
should  include corrections depending on the nuclear $g$ factor. 
Besides a simple one-electron lowest-order nuclear-spin-dependent 
contribution, one should also calculate the second-order corrections 
caused by the hyperfine interaction and the interaction with the 
external magnetic field, taking into account the presence of the 
closed $(1s)^2$ electron shell. 
We perform such calculations in the range $Z=6-32$, where the $2s$ HFS
splitting can be comparable with the Zeeman splitting if the
magnitude of the homogeneous magnetic field is in the range under
consideration. The calculations are based 
on perturbation theory in the parameter $1/Z$ within a rigorous QED
approach. The contributions of zeroth and first orders in $1/Z$ 
are taken into account for the magnetic-dipole correction and 
the contribution of zeroth order is taken into consideration 
for the electric-quadrupole correction.
Also, the $B^2$-dependent correction is calculated, 
including the contributions of zeroth and first orders in $1/Z$.
The obtained results are combined with 
other corrections to get accurate theoretical predictions
for the Breit -- Rabi formula coefficients for lithiumlike ions 
with nonzero nuclear spin.
These predictions will be important for 
experimental investigations that are anticipated in
the near future at University of Mainz and GSI \cite{qui01}.

Relativistic units ($\hbar=c=1$) and the Heaviside charge unit 
($\alpha=e^{2}/4\pi,\,e<0$) are used in the paper. 
In some important cases, the final
formulas contain $\hbar$ and $c$ explicitly to be applicable for
arbitrary system of units.

\section{The Breit -- Rabi formula in the lowest-order  
one-electron approximation}

We consider a lithiumlike ion with nonzero nuclear spin $I$ in a state
of the valent electron with the total electron angular momentum
$j=1/2$. The ion is placed in a homogeneous magnetic field 
$\vec{B}$ directed along the $z$ axis. 
The magnetic splitting is linear with respect to
$\vecb$ only if one of the following conditions is fulfilled:
either $\De E_{\rm mag}\ll \dhfs$ or $\De E_{\rm mag}\gg \dhfs$,
where $\dhfs=E(F+1)-E(F),\ E(F)=E_{n\ka} +\varepsilon_{\rm
hfs}(F)$, $F=I\pm 1/2$ is the total atomic angular momentum, and
$\varepsilon_{\rm hfs}(F)$ is the hyperfine-structure
shift of the valent electron Dirac state with the one-electron energy 
in the Coulomb field of the nucleus
\begin{equation}\label{en}
 E_{n\ka}=\frac{\ga+n_r}{N}\me\,.
\end{equation}
Here $n$ is the principal quantum number,
$\ka=(-1)^{j+l+\frac{1}{2}}(j+\frac{1}{2})$, $l=j\pm \frac{1}{2}$
defines the parity of the state, $n_r=n-|\ka|$ is the radial
quantum number, $\ga=\sqrt{\ka^2-(\az)^2}$,
$N=\sqrt{n_r^2+2n_r\ga+\ka^2}$, and $m_e$ is the electron mass. It
should be emphasized that in case the second inequality is
fulfilled $\De E_{\rm mag}$ must be much less than the distance to
other Dirac's levels. In the intermediate $\vecb$ case, $\De
E_{\rm mag}\sim \dhfs$, we must take into account mixing the HFS
sublevels with the same $\mf$, where $\mf=-F, -F+1, ..., F-1, F$
is the $z$ projection of the total angular momentum. For the
states with the total electron angular momentum $j=1/2$, 
there are only two HFS levels $F=I-1/2$ and
$\Ft=I+1/2$ with the same $\mf=-I+1/2, ...,I-1/2$. This greatly
simplifies the theory.  In what follows, we restrict our 
consideration to the ground state of the valent electron. 
Denoting~\footnote
{In the present paper,
the energy of a Zeeman sublevel $\De E_{\rm mag}$ is counted with
respect to the mean energy $\frac{E^{(2s)}(F)+E^{(2s)}(F')}{2}$ 
of the \hfs\ doublet \cite{bet57, zap79}. To count the energy
from the hyperfine centroid of the doublet
\cite{BR31, HKopf}, one should use the relation
\begin{equation}
\De E_{\rm mag}^{\rm hc}= 
\De E_{\rm mag}-\frac{\dhfs^{(2s)}}{2(2I+1)}\,.
\notag
\end{equation}
}~$\De E_{\rm mag}= E-\frac{E^{(2s)}(F)+E^{(2s)}(F+1)}{2}$, 
one can derive for the Zeeman splitting
\begin{equation}\label{BR1}
\De E_{\rm mag}(x)=\dhfs^{(2s)}\left[a_1\mf x
        \pm
\frac{1}{2} \sqrt{1 +\frac{4\mf}{2I+1}c_1 x +c_2 x^2 }\,\,\right]\,,
\end{equation}
where $x=\mub B/\dhfs^{(2s)}$, $\mub=|e|\hbar/(2m_e c)$ 
is the Bohr magneton,
\begin{align}\label{smallcoeff}
a_1&=-\git \,,   \\
c_1&= g_j +\git\,, \\
c_2&=(g_j +\git)^2\,,
\end{align}
$g_j $ is the ground-state bound-electron $g$ factor of the 
lithiumlike ion,
\begin{equation}\label{gj}
g_j = g_{\rm D}+\Delta g_{\rm int}+\Delta g_{\rm QED}+\Delta g_{\rm rec}^{(e)}+
\Delta g_{\rm NS}+\Delta g_{\rm NP}\,,
\end{equation}
$g_{\rm D} $ is the one-electron Dirac value for a point-charge nucleus,
\begin{eqnarray}\label{gD}
g_{\rm D}=\frac{2[\sqrt{2+2\ga}+1]}{3} = 2 - \frac{(\az)^2}{6}+...\,,
\end{eqnarray}
$\ga=\sqrt{1-(\az)^2}$,
$\Delta g_{\rm int}$ is the interelectronic-interaction 
correction, $\Delta g_{\rm QED}$ is the QED correction, $\Delta g_{\rm
rec}^{(e)}$ is the nuclear recoil correction to the bound-electron
$g$ factor, $\Delta g_{\rm NS}$ is the nuclear size correction,
$\Delta g_{\rm NP}$ is the nuclear polarization correction, $\git$
is the nuclear $g$ factor expressed in the Bohr magnetons,
\begin{equation}\label{defgit}
\git=\frac{\me}{\mpr}(g_I+\Delta g_{\rm rec}^{(n)})\,,
\end{equation}
$\mpr$ is the proton mass, $g_I=\mu/(\mun I)$ , $\mu =\langle
II|\mu_z|II\rangle$ is the nuclear magnetic moment, $\mu_z$ is
the $z$ projection of the nuclear magnetic moment operator
$\vec{\mu}$ acting in the space of nuclear wave functions
$|I\mi\rangle$ with the total angular momentum $I$ and its
projection $\mi$, $\mun=|e|\hbar/(2\mpr c)$ is the nuclear
magneton, and $\Delta g_{\rm rec}^{(n)}$ is the recoil correction
to the bound-nucleus $g$ factor (see section \ref{Numer}). Eq.
(\ref{BR1}) is usually called the Breit -- Rabi formula (see,
e.g., Refs. \cite{BR31, HKopf, bet57, zap79, mosk06}). It covers Zeeman
splitting from weak ($x\ll 1$) to strong ($x\gg 1$) fields
including the intermediate region. For $\Ft=I+\frac{1}{2}$ and 
$\mf=\pm(I+\frac{1}{2})$ the splitting is linear in the first order
of perturbation theory under arbitrary magnetic induction,
\begin{equation}\label{BRl1}
\De E_{\rm mag}(x)=\dhfs^{(2s)}\left[\frac{1}{2}
\pm d_1 x\right] \,,
\end{equation}
where
\begin{equation}
d_1 = \frac{1}{2}g_j-I\git\,
\end{equation}
and the ``$-$'' and ``$+$'' signs
refer to $\mf=-(I+\frac{1}{2})$ and
$\mf=I+\frac{1}{2}$, respectively.

For Li-like ions with $I=1/2$, $F=0$ and $\Ft=1$ and, therefore,
the two mixed sublevels have $\mf=0$. In this case the Breit --
Rabi formula takes the form
\begin{equation}\label{BR2} \De E_{\rm mag}(x)=
  \pm\frac{\dhfs^{(2s)}}{2} \sqrt{1 +c_2 x^2 } \,,
\end{equation}
and for $\mf=\pm 1$ the effect is described by Eq. (\ref{BRl1})
with $d_1 = \frac{1}{2}(g_j-\git)$.

If the magnetic field is strong, $\Delta E_{\rm mag} \gg\dhfs^{(2s)}$,
Eqs. (\ref{BR1}), (\ref{BRl1}), and (\ref{BR2}) convert into
formulas for the anomalous Zeeman effect of the $2s$ state. 

In case the energy-level shift
(splitting) due to interaction with $\vec{B}$ is much smaller than
the hyperfine-structure splitting, $\Delta E_{\rm mag}\ll \dhfs^{(2s)}$,
we can express the linear-dependent part of this shift in terms of
the atomic $g$ factor,
\begin{equation}\label{magsplit}
\Delta E_{\rm mag}= \pm\frac{\dhfs^{(2s)}}{2}+g(F)\mub B\mf \,,
\end{equation}
where, to the lowest-order approximation (see, e.g., Ref. \cite{bet57}),
\begin{equation}\label{gat}
g(F)= g_{\rm D}Y_{\rm el}(F)-
\frac{\me}{\mpr}g_I Y_{\rm nuc}^{(\mu)}(F)\,,
\end{equation}
\begin{eqnarray} \label{Yel}
Y_{\rm el}(F)=\frac{F(F+1)+3/4-I(I+1)}{2F(F+1)}=
\begin{cases}
  -\frac{1}{2I+1}   &\text{for $F=I-\frac{1}{2}$}\\
  \frac{1}{2I+1}        &\text{for $F=I+\frac{1}{2}$}
\end{cases}\,,
\end{eqnarray}
\begin{eqnarray} \label{Ynucdip}
Y_{\rm nuc}^{(\mu)}(F)=\frac{F(F+1)+I(I+1)-3/4}{2F(F+1)}=
\begin{cases}
  \frac{2(I+1)}{2I+1}   &\text{for $F=I-\frac{1}{2}$}\\
  \frac{2I}{2I+1}        &\text{for $F=I+\frac{1}{2}$}
\end{cases}\,.
\end{eqnarray}

The total one-electron $2s$ $g$-factor value of a Li-like ion
with nonzero nuclear spin can be represented by
\begin{eqnarray}\label{gtot}
g(F)&=&(g_{\rm D}+\Delta g_{\rm int}+
\Delta g_{\rm QED}+\Delta g_{\rm rec}^{(e)}+
\Delta g_{\rm NS}+\Delta g_{\rm NP}) Y_{\rm el}(F)
 \nonumber \\
&&-\frac{\me}{\mpr}(g_I+\Delta g_{\rm rec}^{(n)}) Y_{\rm
nuc}^{(\mu)}(F)+\delta g_{\rm HFS}^{(2s)}(F) \,,
\end{eqnarray}
where the HFS correction $\delta g_{\rm HFS}^{(2s)}(F)= \delta
g_{\rm HFS(\mu)}^{(2s)}(F)+\delta g_{\rm HFS(Q)}^{(2s)}(F)$  
\cite{mosk07os} is briefly discussed below.

\section{Hyperfine-interaction corrections 
to the ground state $g$ factor}\label{HFScorr}

Let us start our consideration of the HFS correction 
to the ground-state $g$ factor of a Li-like ion with
the one-electron approximation. In this approximation,
the interaction of the ion with the magnetic field 
can be represented as
\begin{equation}
V_{\vecb} = V_{\vecb}^{(e)} + V_{\vecb}^{(n)}\,.
\end{equation}
Here $V_{\vecb}^{(e)}$ describes the interaction of the valent $2s$
electron with the homogeneous magnetic field,
\begin{equation}\label{W}
V_{\vecb}^{(e)}=-e(\vec{\alpha}\cdot
\vec{A})=\frac{|e|}{2}(\vec{\alpha}\cdot [\vecb\times\vec{r}])\,,
\end{equation}
where the vector $\vec{\alpha}$ incorporates the Dirac $\alpha$
matrices, and
\begin{equation}\label{Wn}
V_{\vecb}^{(n)}=-(\vec{\mu}\cdot \vecb)
\end{equation}
describes the interaction of the nuclear magnetic moment
$\vec{\mu}$ with $\vecb$. 
The hyperfine-interaction operator is given by the sum
\begin{equation}
V_{\rm HFS}=V_{\rm HFS}^{(\mu)}+V_{\rm HFS}^{(Q)}\,,
\end{equation}
where $V_{\rm HFS}^{(\mu)}$ and $V_{\rm HFS}^{(Q)}$ are the
magnetic-dipole and electric-quadrupole hyperfine-interaction
operators, respectively. In the point-dipole approximation,
\begin{equation}\label{FB}
V_{\rm HFS}^{(\mu)}=
\frac{|e|}{4\pi}\frac{(\vec{\alpha}\cdot[\vec{\mu}\times
\vec{r}])}{r^3}\,,
\end{equation}
and, in the point-quadrupole approximation,
\begin{equation}
V_{\rm HFS}^{(Q)} = -\alpha \sum_{m=-2}^{m=2}Q_{2m}
\eta_{2m}^*(\vec{n})\,.
\end{equation}
Here $ Q_{2m}=\sum_{i=1}^Zr_i^2C_{2m}(\vec{n}_i)$ is the operator
of the electric-quadrupole moment of the nucleus,
$\eta_{2m}=C_{2m}(\vec{n})/r^3 $ is an operator that acts on
electron variables, $\vec{n}=\vec{r}/r$, $\vec{n}_i=\vec{r}_i/r_i$, 
$\vec{r}$ is the
position vector of the electron, $\vec{r}_i$ is the position
vector of the $i$-th proton in the nucleus, $C_{lm}=
\sqrt{4\pi/(2l+1)}\,Y_{lm}$, and $Y_{lm}$ is a spherical harmonic.
It must be stressed that the electric-quadrupole interaction
should be taken into account only for ions with $I > 1/2$.

An unperturbed atomic eigenstate that corresponds 
to given values of $F$ and $\mf$ is  
a linear combination of
products of electron and nuclear wave functions,
\begin{equation}\label{ket}
|nljIF\mf\rangle=\sum_{\mj, \mi}C_{j\mj
I\mi}^{F\mf}|nlj\mj\rangle|I\mi\rangle.
\end{equation}
Here $C_{j\mj I\mi}^{F\mf}$ are the Clebsch-Gordan coefficients,
$|nlj\mj\rangle$ are the unperturbed one-electron wave functions, 
which are four-component eigenvectors of the Dirac equation for
the Coulomb field, with the total angular momentum $j$ and its
projection $\mj$.

In the one-electron approximation, the magnetic-dipole and 
electric-quadrupole hyperfine-interaction corrections 
to the ground-state $g$ factor of the Li-like ion are given by
\begin{equation}\label{HFScg}
\delta g_{\rm HFS(\mu, Q)}^{{\rm one-el.}(2s)}=\frac{2}{\mub B\mf}
\sum_{\mj\mi}\sum_{\mj'\mi'}C^{F\mf}_{\frac{1}{2}\mj I\mi}
C^{F\mf}_{\frac{1}{2}\mj' I\mi'}
\la I\mi|\sum_n^{(\eps_n\neq\eps_v)}
\frac{\la v|V_{\vecb}^{(e)}|n\ra\la n|V_{\rm
HFS}^{(\mu, Q)}|v'\ra}{\veps_v-\veps_n}|I\mi'\ra\,,
\end{equation}
where $|v\ra=|20\frac{1}{2}\mj\rangle$ and $|v'\ra=|20\frac{1}{2}\mj'\rangle$ 
are the $2s$ states of the valent electron with the
angular momentum projections $\mj$ and $\mj'$, respectively, 
$|n\ra\equiv |nlj\mj\ra$, $\veps_v = E_{2,-1}$, and 
$\veps_n\equiv E_{n\ka}$.
The summation in (\ref{HFScg}) runs over
discrete as well as continuum states. 
The corresponding diagrams are presented in Fig. \ref{Diagrams2}.

The total hyperfine-interaction correction 
to the ground-state $g$ factor of the Li-like ion is given by
\begin{equation}\label{gcorrtot}
\delta g_{\rm HFS}^{(2s)}=\delta g_{\rm{HFS}(\mu)}^{(2s)}+ \delta
g_{\rm{HFS}(Q)}^{(2s)}
\end{equation}
with
\begin{equation}\label{gcorrdip}
\delta g_{\rm{HFS}(\mu)}^{(2s)}=\alpha^2
Z\frac{1}{12}\frac{\mu}{\mun}\frac{\me}{\mpr} \frac{1}{I}Y_{\rm
nuc}^{(\mu)}(F)
[S_2(\az) + \frac{1}{Z} B_{\mu}(\az) + 
\frac{1}{Z^2} C_{\mu}(\az)+\dots]
\end{equation} 
and
\begin{equation}\label{gcorrquadr}
\delta g_{\rm{HFS}(Q)}^{(2s)}=\alpha^4 Z^3\frac{23}{2160}
Q\left(\frac{\me c}{\hbar}\right)^2 Y_{\rm nuc}^{(Q)}(F) 
[T_2(\az) + \frac{1}{Z} B_{Q}(\az) + 
\frac{1}{Z^2} C_{Q}(\az)+\dots]\,.
\end{equation}
Here the angular factor is
\begin{eqnarray} \label{Ynucquadr}
Y_{\rm nuc}^{(Q)}(F)=
\begin{cases}
  -\frac{(I+1)(2I+3)}{I(2I-1)(2I+1)}   &\text{for $F=I-\frac{1}{2}$}\\
  \,\,\,\,\,\,    \frac{1}{2I+1}        &\text{for $F=I+\frac{1}{2}$}
\end{cases}\,,
\end{eqnarray}
and $Q=2\langle II|Q_{20}|II \rangle$ is the electric-quadrupole
moment of the nucleus. 
The functions
\begin{equation}\label{defS2}
S_2(\az)=\frac{12}{\alpha^2 Z\,\frac{\me}{\mpr} g_I
Y_{\rm nuc}^{(\mu)}(F)}\,\delta g_{\rm HFS(\mu)}^{{\rm one-el.}(2s)}
\end{equation}
and
\begin{equation}\label{defT2}
T_2(\az)=\frac{2160}{23 \alpha^4 Z^3\, Q
\left(\frac{\me c}{\hbar}\right)^2
Y_{\rm nuc}^{(Q)}(F)}\,\delta g_{\rm HFS(Q)}^{{\rm one-el.}(2s)}
\end{equation}
determine the one-electron contributions, which are discussed in 
detail in Ref. \cite{mosk04}.
For the point-charge nucleus, the functions $S_2(\az)$ and $T_2(\az)$ 
are~\cite{mosk04, mosk07os}
\begin{align}\label{S2}
S_2(\az)&=\frac{8}{3N}
\biggl\{\frac{1}{N+2}\biggl[N + \frac{10(N+1)}{3N}\biggr]
+\frac{(\az)^2}{\ga(\ga+1)}\biggl[\frac{2(N+1)}{3-4(\az)^2} + 1\biggr]
-\frac{1}{\ga}\biggr\}
\notag \\
&=1+\frac{229}{144}(\az)^2+\dots
\end{align}
and
\begin{align}
T_2(\az)&=\frac{192[(N+\ga+1) \{18 + 24\ga - 12N + 8\ga
N^2\} + 15(1+\ga)]} {23\ga N^3[15 - 16(\az)^2](N + \ga+1)^2} 
\notag \\
&=1+\frac{427}{276}(\az)^2+\dots\,,
\end{align}
where 
$N=\sqrt{2(1+\ga)}$.

The interelectronic-interaction correction $B_\mu(\az)$ 
can be calculated within the rigorous QED approach \cite{mosk07os}. 
The interaction of the electrons 
with the Coulomb field of the nucleus is included in the unperturbed
Hamiltonian, i.e. the Furry picture is used. The perturbation theory 
is formulated with the technique of the two-time Green function (TTGF) 
\cite{sha02c,shaVUZ}. To simplify the calculations, the closed
${(1s)^2}$ shell is regarded as belonging to a redefined vacuum. 
With this vacuum, the Fourier transform of TTGF can be introduced by
\begin{equation}
\begin{split}
{\cal{G}}(E;{\vec{x'}};{\vec{x}})&\delta(E-E')=\frac{1}{2\pi
i}\int\limits_{-\infty}\limits^{\infty}dx^{0}dx'^{0}
\exp(iE'x'^{0}-iEx^{0}) \\&
\times\langle0_{(1s)^2}|T\psi(x'^{0},{\vec{x'}})\psi^{\dag}(x^{0},
{\vec{x}})|0_{(1s)^2}\rangle,
\end{split}
\end{equation}
where $\psi(x^{0},{\vec{x}})$ is the electron-positron field
operator in the Heisenberg representation and $T$ is the
time-ordered product operator. The energy shift of a state $a$ can be 
expressed in terms of the TTGF defined by
\begin{equation}\label{Gaa}
g_{aa}(E) = \langle u_{a}|{\cal{G}}(E)| u_{a}\rangle \equiv \int
{d\vec{x}}d{\vec{x'}} u_{a}^{\dag}({\vec{x'}}){\cal{G}}(E;{\vec
{x'}};{\vec{x}}) u_{a}({\vec{x}}),
\end{equation}
where $u_{a}({\vec{x}})$ is the unperturbed Dirac wave function of the
state $a$.
Using the Sz.-Nagy and Kato technique \cite{SNK}, one can derive for the 
total energy shift $\De E_a\equiv E_a - E_a^{(0)}$ \cite{sha02c,shaVUZ}
\begin{equation}\label{DeEtot}
\De E_{a}=\frac{\frac{1}{2\pi i}\oint\limits_{\Gamma}{dE
\,\De E \De g_{aa}(E)}}
{1 + \frac{1}{2\pi i}\oint\limits_{\Gamma}{dE
\, \De g_{aa}(E)}}\,,
\end{equation}
where $\De E \equiv  E-E_a^{(0)}$, 
$\De g_{aa}(E)\equiv  g_{aa}(E)- g^{(0)}_{aa}(E)$, and 
$g^{(0)}_{aa}(E)=(E-E_a^{(0)})^{-1}$. The integrals in the
complex $E$-plane are taken along the contour $\Ga$ which
surrounds the pole of $g_{aa}(E)$ corresponding to the 
level $a$ and keeps outside 
all other singularities. The contour $\Gamma$ is oriented
counter-clockwise.

To first three orders of the perturbation theory, the energy
shift is given by
\begin{align} 
\De E_{a}^{(1)}&=\frac{1}{2\pi
i}\oint_{\Gamma}{dE \, \De E\De
g_{aa}^{(1)}(E)}\,,
\\
\label{DeE2}
\De E_{a}^{(2)}&=\frac{1}{2\pi
i}\oint_{\Gamma}{dE \, \De E\De
g_{aa}^{(2)}(E)}   
-\left(\frac{1}{2\pi i}\oint_{\Gamma}{dE \,
\De E \De g_{aa}^{(1)}(E)}\right) \left(\frac{1}{2\pi
i}\oint_{\Gamma}{dE \, \De g_{aa}^{(1)}(E)}\right) \,,
\\
\label{DeE3}
\De E_{a}^{(3)}&=\frac{1}{2\pi
i}\oint_{\Gamma}{dE \, \De E\De
g_{aa}^{(3)}(E)}   
-\left(\frac{1}{2\pi i}\oint_{\Gamma}{dE \,
\De E \De g_{aa}^{(2)}(E)}\right) \left(\frac{1}{2\pi
i}\oint_{\Gamma}{dE \, \De g_{aa}^{(1)}(E)}\right)
                      \nonumber \\
&-\left(\frac{1}{2\pi i}\oint_{\Gamma}{dE \,
\De E \De g_{aa}^{(1)}(E)}\right) \left(\frac{1}{2\pi
i}\oint_{\Gamma}{dE \, \De g_{aa}^{(2)}(E)}\right)
                     \nonumber \\
&+\left(\frac{1}{2\pi i}\oint_{\Gamma}{dE \,
\De E \De g_{aa}^{(1)}(E)}\right) \left(\frac{1}{2\pi
i}\oint_{\Gamma}{dE \, \De g_{aa}^{(1)}(E)}\right)^2 \,.
\end{align}
The redefinition of the vacuum changes $i0$ to $-i0$ in the
electron propagator denominators corresponding to the closed
$(1s)^2$ shell. In other words it means replacing the standard
Feynman contour of integration over the electron energy $C$ with a
new contour $C'$ (Fig. \ref{E-plane}). The second-order 
contribution is defined by the diagrams presented in Fig. 
\ref{Diagrams2}. Its evaluation according to Eq. (\ref{DeE2}) yields
formula (\ref{HFScg}). In the formalism under consideration, the
lowest-order interelectronic-interaction and the radiative
corrections to Eq. (\ref{HFScg}) are described by the third-order 
diagrams presented in Fig. \ref{Diagrams3} and, according to 
Eq. (\ref{DeE3}), by some products of the low-order diagrams
depicted in Figs. \ref{Diagrams31} and \ref{Diagrams32}. 
According to Fig. \ref{E-plane}, to separate 
the interelectronic-interaction corrections, the contour $C'$ must be 
divided into two parts, $C$ and $C_{\rm int}$.
The integral along the standard
Feynman contour $C$ gives the one-electron radiative correction.
The integral along the contour $C_{\rm int}$ describes the interaction
of the valent electron with the closed shell electrons. 
Formula (\ref{DeE3}) allows one to evaluate the 
interelectronic-interaction correction $B_{\mu}(\az)$ \cite{mosk07os}.
The results of this evaluation will be presented in the next section
together with other related corrections to the Breit~-- Rabi formula.

\section{Corrections to the Breit -- Rabi formula for the ground state}

Now we assume that the Zeeman splitting $\De E_{\rm mag}$ of the $2s$ HFS
levels $F=I-1/2$ and $\Ft=I+1/2$ is much smaller than the distance
to other levels but is comparable with $\dhfs^{(2s)}$. The
unperturbed eigenstates form a two-dimensional subspace
$\Omega=\{|1^{(0)}\ra,\ |2^{(0)}\ra\}$, where 
$|1^{(0)}\ra=|20\frac{1}{2}IF\mf\ra,\
|2^{(0)}\ra=|20\frac{1}{2}I\Ft\mf\ra$. 
Employing the perturbation theory for degenerate states \cite{sha02c} 
with energy $\veps_v$ we denote the projector on $\Omega$ by
\begin{equation}
P^{(0)} = \sum_{i=1}^2 |i^{(0)}\ra \la i^{(0)}|\,.
\end{equation}
We project the Green function ${\cal{G}}(E)$ on the subspace $\Omega$
\begin{equation}
g(E) = P^{(0)}{\cal{G}}(E)P^{(0)}\,,
\end{equation}
where, as in Eq. (\ref{Gaa}), 
the integration over the electron coordinates is implicit. In this
case we can choose a contour $\Gamma$ in the complex $E$-plane in
a way that it surrounds all $g(E)$ poles, which correspond
to the states under consideration, and keeps outside all other 
singularities of $g(E)$.
As in the case of a single level, to the zeroth-order approximation
one easily finds
\begin{equation}
g^{(0)}(E)=\sum_{i=1}^2\frac{|i^{(0)}\ra \la i^{(0)}|}{E-E_i^{(0)}}\,.
\end{equation}
We introduce the operators $K$ and $P$ by
\begin{align}
\label{K}
K&\equiv \frac{1}{2\pi
i}\oint_{\Gamma}{dE \, E g(E)}\,, \\
\label{P}
P&\equiv\frac{1}{2\pi
i}\oint_{\Gamma}{dE \, g(E)}\,.
\end{align}
As it is shown in Ref. \cite{sha02c}, the energy levels 
are determined from the equation
\begin{equation}
{\rm det}(H-E)=0\,,
\end{equation}
where 
\begin{equation}
H=P^{-\frac{1}{2}}KP^{-\frac{1}{2}}\,.
\end{equation}
The operators $K$ and $P$ are constructed by formulas
(\ref{K}) and (\ref{P}) 
\begin{align}
K=&K^{(0)}+K^{(1)}+K^{(2)}+K^{(3)}+\dots\,, \\
P=&P^{(0)}+P^{(1)}+P^{(2)}+P^{(3)}+\dots\,,
\end{align}
where the superscript indicates the order of the perturbation
theory in a small parameter.
The operator $H$ is
\begin{equation}\label{H}
H=H^{(0)}+H^{(1)}+H^{(2)}+H^{(3)}+\dots\,,
\end{equation}
where
\begin{align}
H^{(0)}&=K^{(0)}\,, \\
H^{(1)}&=K^{(1)}-\frac{1}{2}P^{(1)}K^{(0)}-\frac{1}{2}K^{(0)}P^{(1)}\,, \\
H^{(2)}&=K^{(2)}-\frac{1}{2}P^{(2)}K^{(0)}-\frac{1}{2}K^{(0)}P^{(2)}
-\frac{1}{2}P^{(1)}K^{(1)}-\frac{1}{2}K^{(1)}P^{(1)}  \notag \\
&+\frac{3}{8}P^{(1)}P^{(1)}K^{(0)}+\frac{3}{8}K^{(0)}P^{(1)}P^{(1)}
+\frac{1}{4}P^{(1)}K^{(0)}P^{(1)}\,,   \\
H^{(3)}&=K^{(3)}-\frac{1}{2}P^{(3)}K^{(0)}-\frac{1}{2}K^{(0)}P^{(3)}
-\frac{1}{2}P^{(1)}K^{(2)}-\frac{1}{2}K^{(2)}P^{(1)}
-\frac{1}{2}P^{(2)}K^{(1)}-\frac{1}{2}K^{(1)}P^{(2)}  \notag \\
&+\frac{3}{8}P^{(1)}P^{(2)}K^{(0)}+\frac{3}{8}K^{(0)}P^{(1)}P^{(2)}
+\frac{3}{8}P^{(2)}P^{(1)}K^{(0)}+\frac{3}{8}K^{(0)}P^{(2)}P^{(1)} 
\notag \\
&+\frac{1}{4}P^{(1)}K^{(0)}P^{(2)}+\frac{1}{4}P^{(2)}K^{(0)}P^{(1)}
+\frac{3}{8}P^{(1)}P^{(1)}K^{(1)}+\frac{3}{8}K^{(1)}P^{(1)}P^{(1)}
+\frac{1}{4}P^{(1)}K^{(1)}P^{(1)}
\notag \\
&-\frac{5}{16}P^{(1)}P^{(1)}P^{(1)}K^{(0)}
-\frac{5}{16}K^{(0)}P^{(1)}P^{(1)}P^{(1)}
\notag \\
&-\frac{3}{16}P^{(1)}P^{(1)}K^{(0)}P^{(1)}
-\frac{3}{16}P^{(1)}K^{(0)}P^{(1)}P^{(1)}\,.
\end{align}

Taking into account only the relevant contributions 
of kind $\alpha\times\mu/\mun\times B$ and 
$\alpha\times B\times B$, where $\alpha$ comes from the interelectronic 
interaction, we obtain for the third-order term in Eq. (\ref{H})
\begin{align}\label{DeH3}
H_{jk}^{(3)}&=\frac{1}{2\pi i}
\oint_{\Gamma}{dE \, \De E\De g_{jk}^{(3)}(E)}   
-\frac{1}{2}\sum_{l=1}^2
\biggl[
\left(\frac{1}{2\pi i}\oint_{\Gamma}{dE \, 
\De g_{jl}^{(1)}(E)}\right)
\left(\frac{1}{2\pi i}\oint_{\Gamma}{dE \,
\De E \De g_{lk}^{(2)}(E)}\right)
\nonumber \\
&+\left(\frac{1}{2\pi i}\oint_{\Gamma}{dE \,
\De E \De g_{jl}^{(2)}(E)}\right) 
\left(\frac{1}{2\pi i}\oint_{\Gamma}{dE \, 
\De g_{lk}^{(1)}(E)}\right)
\biggr]
\nonumber \\
&-\frac{1}{2}\sum_{l=1}^2
\biggl[
\left(\frac{1}{2\pi i}\oint_{\Gamma}{dE \, 
\De g_{jl}^{(2)}(E)}\right)
\left(\frac{1}{2\pi i}\oint_{\Gamma}{dE \,
\De E \De g_{lk}^{(1)}(E)}\right)
\nonumber \\
&+\left(\frac{1}{2\pi i}\oint_{\Gamma}{dE \,
\De E \De g_{jl}^{(1)}(E)}\right) 
\left(\frac{1}{2\pi i}\oint_{\Gamma}{dE \, 
\De g_{lk}^{(2)}(E)}\right)
\biggr]\,, 
\end{align}
where $\De E \equiv  E-\veps_v$, $j,k=1,2$.

Keeping only the three lowest-order terms in $B$, 
we get the following equation for the perturbed energies:
\begin{equation}\label{secur}
\begin{vmatrix}
h_0(F)+h_1(F)B+h_2(F)B^2-E &
\tilde{h}_1(F,\Ft)B+\tilde{h}_2(F,\Ft)B^2   \\
\tilde{h}_1(\Ft,F)B+\tilde{h}_2(\Ft,F)B^2
&h_0(\Ft)+h_1(\Ft)B+h_2(\Ft)B^2-E
\end{vmatrix}
=0 \,.
\end{equation}
Here $F=I-\frac{1}{2}$, $\Ft=I+\frac{1}{2}$,
\begin{equation}
h_0(k) = E(k)
\end{equation}
is the energy of the HFS level,
\begin{align}
h_1(k) &= \frac{1}{B}[\De E_{(\vecb)}^{(1)}(k,k)+
\De E_{(\mu)}^{(2)}(k,k)   + \De E_{(Q)}^{(2)}(k,k) + 
\De E_{(\mu)}^{(3)}(k,k)   + \De E_{(Q)}^{(3)}(k,k) ]\notag   \\ 
&+ (\Delta g_{\rm int}+
\Delta g_{\rm QED}+\Delta
g_{\rm rec}^{(e)}+ \Delta g_{\rm NS}+\Delta g_{\rm NP}) Y_{\rm
el}(k)\mub \mf -
 \Delta g_{\rm rec}^{(n)}Y_{\rm nuc}^{(\mu)}(k)\mun \mf \\
&= g(k)\mub \mf\,, \notag
\end{align}
\begin{equation}\label{h2k}
h_2(k)=\frac{1}{B^2}[\De E_{(\vecb)}^{(2)}(k,k) +
\De E_{(\vecb)}^{(3)}(k,k)]\,,
\end{equation}
\begin{align}\label{h1tjk}
\tilde{h}_1(j,k)&=\frac{1}{B}[\De E_{(\vecb)}^{(1)}(j,k)+
\De E_{(\mu)}^{(2)}(j,k)   + \De E_{(Q)}^{(2)}(j,k) + 
\De E_{(\mu)}^{(3)}(j,k)   + \De E_{(Q)}^{(3)}(j,k) ]\notag \\ 
&+(\Delta_{\rm int}+\Delta _{\rm QED}+\Delta _{\rm rec}^{(e)}+ 
\Delta _{\rm NS}+\Delta _{\rm NP})\mub 
-\Delta _{\rm rec}^{(n)}\mun \,,
\end{align}
\begin{equation}\label{h2tjk}
\tilde{h}_2(j,k)=\frac{1}{B^2}[\De E_{(\vecb)}^{(2)}(j,k) +
\De E_{(\vecb)}^{(3)}(j,k)]\,,
\end{equation}
where $j,\,k=F,\,\Ft$. $\Delta _{\rm int}$, 
$\Delta _{\rm QED}$, $\Delta _{\rm rec}^{(e)}$, $\Delta_{\rm NS}$, 
and $\Delta _{\rm NP}$ are the interelectronic-interaction, 
QED, nuclear recoil,
nuclear size, and nuclear polarization corrections. They are
similar to the corresponding corrections to $ h_1(k)$ but have
a different angular factor as well as $\Delta _{\rm rec}^{(n)}$. 
It should be noted that we have
neglected here terms describing virtual transitions into excited
nuclear states via the direct interaction of the nucleus with the
magnetic field \cite{jent06}. 
The energy shifts are
\begin{align}
\label{DeE1CC}
\De E_{(\vecb)}^{(1)}(j,k)&=
\sum_{\mj\mi}\sum_{\mj'\mi'}C^{j\mf}_{\frac{1}{2}\mj I\mi}
C^{k\mf}_{\frac{1}{2}\mj' I\mi'}
\la I\mi|V_{\vecb}|I\mi'\ra\,,            \\
\label{DeE2CC}
\De E_{(\mu,Q,\vecb)}^{(2)}(j,k)&=
\sum_{\mj\mi}\sum_{\mj'\mi'}C^{j\mf}_{\frac{1}{2}\mj I\mi}
C^{k\mf}_{\frac{1}{2}\mj' I\mi'}
\la I\mi|I_{\mu,Q,\vecb}^{(2)}|I\mi'\ra\,,  \\
\label{DeE3CC} 
\De E_{(\mu,Q,\vecb)}^{(3)}(j,k)&=
\sum_{\mj\mi}\sum_{\mj'\mi'}C^{j\mf}_{\frac{1}{2}\mj I\mi}
C^{k\mf}_{\frac{1}{2}\mj' I\mi'}
\la I\mi|I_{\mu,Q,\vecb}^{(3a)}+I_{\mu,Q,\vecb}^{(3b)}
+I_{\mu,Q,\vecb}^{(3c)}+I_{\mu,Q,\vecb}^{(3d)} 
|I\mi'\ra \,,
\end{align}
where
\begin{equation}\label{I}
I_{\mu,Q,\vecb}^{(2)}=2f
\sum_n^{(\eps_n\neq\eps_v)}
\frac{\la v|V_{\vecb}^{(e)}|n\ra\la n|W|v'\ra}{\veps_v-\veps_n}\,,
\end{equation}
\begin{align}
\label{Ia}
I_{\mu,Q,\vecb}^{(3a)}&= f\sum_{\veps_c=E_{1,-1}} 
\biggl(
\sum_{n_1,n_2}^{(\veps_{n_1}\neq \veps_v,\,\veps_{n_2}\neq \veps_v)}
\frac{2}{(\veps_v-\veps_{n_1})(\veps_v-\veps_{n_2})}
\biggl[
\la v|V_{\vecb}^{(e)}|n_1\ra\la n_1|W|n_2\ra
\la n_2 c|I(0)|v' c\ra  
 \nonumber \\
&+\la v|V_{\vecb}^{(e)}|n_1\ra\la n_1 c|I(0)|n_2 c\ra
\la n_2|W|v'\ra
 +\la v|W|n_1\ra\la n_1|V_{\vecb}^{(e)}|n_2\ra
\la n_2 c|I(0)|v' c\ra
\biggr]
\nonumber \\
&-\sum_{\veps_{\tv}=\veps_v}\sum_{n}^{(\veps_{n}\neq \veps_v)}
\frac{2}{(\veps_v-\veps_{n})^2}
\biggl[
\la v|V_{\vecb}^{(e)}|n\ra\la n|W|\tv\ra
\la \tv c|I(0)|v' c\ra
\nonumber \\
&+\la v|V_{\vecb}^{(e)}|n\ra\la n c|I(0)|\tv c\ra
\la \tv|W|v'\ra
+\la v|V_{\vecb}^{(e)}|\tv\ra\la \tv|W|n\ra
\la n c|I(0)|v' c\ra
\biggr]
 \biggr)\,,
\\
\label{Ib}
I_{\mu,Q,\vecb}^{(3b)}&= -f\sum_{\veps_c=E_{1,-1}} 
\biggl(
\sum_{n_1,n_2}^{(\veps_{n_1}\neq \veps_v,\,\veps_{n_2}\neq \veps_v)}
\frac{2}{(\veps_v-\veps_{n_1})(\veps_v-\veps_{n_2})}
\biggl[
\la v|V_{\vecb}^{(e)}|n_1\ra\la n_1|W|n_2\ra
\la n_2 c|I(\om)|c v'\ra  
 \nonumber \\
&+\la v|V_{\vecb}^{(e)}|n_1\ra\la n_1 c|I(\om)|c n_2\ra
\la n_2|W|v'\ra
 +\la v|W|n_1\ra\la n_1|V_{\vecb}^{(e)}|n_2\ra
\la n_2 c|I(\om)|c v'\ra
\biggr]
\nonumber \\
&-\sum_{\veps_{\tv}=\veps_v}\sum_{n}^{(\veps_{n}\neq \veps_v)}
\frac{2}{(\veps_v-\veps_{n})^2}
\biggl[
\la v|V_{\vecb}^{(e)}|n\ra\la n|W|\tv\ra
\la \tv c|I(\om)|c v'\ra
\nonumber \\
&+\la v|V_{\vecb}^{(e)}|n\ra\la n c|I(\om)|c \tv\ra
\la \tv|W|v'\ra
+\la v|V_{\vecb}^{(e)}|\tv\ra\la \tv|W|n\ra
\la n c|I(\om)|c v'\ra
\biggr]
\nonumber \\
&+\sum_{\veps_{\tv}=\veps_v}\sum_{n}^{(\veps_{n}\neq \veps_v)}
\frac{2}{\veps_v-\veps_{n}}
\biggl[
\la v|V_{\vecb}^{(e)}|n\ra\la n|W|\tv\ra
\la \tv c|I'(\om)|c v'\ra
\nonumber \\
&+\la v|V_{\vecb}^{(e)}|n\ra\la n c|I'(\om)|c \tv\ra
\la \tv|W|v'\ra
+\la v|V_{\vecb}^{(e)}|\tv\ra\la \tv|W|n\ra
\la n c|I'(\om)|c v'\ra
\biggr]
\nonumber \\
&+\sum_{\veps_{\tv}=\veps_v}\sum_{\veps_{\bv}=\veps_v}
\la v|V_{\vecb}^{(e)}|\tv\ra\la \tv c|I''(\om)|c \bv\ra
\la \bv|W|v'\ra 
\biggr)\,,
\\
\label{Ic}
I_{\mu,Q,\vecb}^{(3c)}&= f\sum_{\veps_c=E_{1,-1}} 
\biggl(
\sum_{n_1,n_2}^{(\veps_{n_1}\neq \veps_v,\,\veps_{n_2}\neq \veps_c)}
\frac{2}{(\veps_v-\veps_{n_1})(\veps_c-\veps_{n_2})}
\biggl[
\la v|V_{\vecb}^{(e)}|n_1\ra\la c|W|n_2\ra
\la n_1 n_2|I(0)|v' c\ra  
\nonumber \\
&+\la v|V_{\vecb}^{(e)}|n_1\ra\la n_1 c|I(0)|v' n_2\ra
\la n_2|W|c\ra
+\la v|W|n_1\ra\la c|V_{\vecb}^{(e)}|n_2\ra
\la n_1 n_2|I(0)|v' c\ra
\nonumber \\  
&+\la v|W|n_1\ra\la n_1 c|I(0)|v' n_2\ra
\la n_2|V_{\vecb}^{(e)}|c\ra
\biggr]
\nonumber \\
&+\sum_{n_1,n_2}^{(\veps_{n_1}\neq \veps_c,\,\veps_{n_2}\neq \veps_c)}
\frac{2}{(\veps_c-\veps_{n_1})(\veps_c-\veps_{n_2})}
\biggl[
\la c|V_{\vecb}^{(e)}|n_1\ra
\la n_1|W|n_2\ra\la n_2 v|I(0)|c v'\ra
\nonumber \\
&+\la c|V_{\vecb}^{(e)}|n_1\ra\la n_1 v|I(0)|n_2 v'\ra 
\la n_2|W|c\ra 
+\la c|W|n_1\ra\la n_1|V_{\vecb}^{(e)}|n_2\ra
\la n_2 v|I(0)|c v'\ra
\biggr]
\nonumber \\
&-\sum_{\veps_{\tc}=E_{1,-1}}\sum_{n}^{(\veps_{n}\neq \veps_c)}
\frac{2}{(\veps_c-\veps_{n})^2}
\biggl[
\la c|V_{\vecb}^{(e)}|n\ra\la n|W|\tc\ra
\la \tc v|I(0)|c v'\ra
\nonumber \\
&+\la c|V_{\vecb}^{(e)}|n\ra\la n v|I(0)|\tc v'\ra
\la \tc|W|c \ra
+\la c|V_{\vecb}^{(e)}|\tc\ra\la \tc|W|n \ra
\la n v|I(0)|c v'\ra
\biggr]
 \biggr)\,,
\end{align}
\begin{align}\label{Id}
I_{\mu,Q,\vecb}^{(3d)}&= -f\sum_{\veps_c=E_{1,-1}} 
\biggl(
\sum_{n_1,n_2}^{(\veps_{n_1}\neq \veps_v,\,\veps_{n_2}\neq \veps_c)}
\frac{2}{(\veps_v-\veps_{n_1})(\veps_c-\veps_{n_2})}
\biggl[
\la v|V_{\vecb}^{(e)}|n_1\ra\la c|W|n_2\ra
\la n_1 n_2|I(\om)|c v'\ra 
\nonumber \\
&+\la v|V_{\vecb}^{(e)}|n_1\ra\la n_1 c|I(\om)|n_2 v'\ra 
\la n_2|W|c\ra
+\la v|W|n_1\ra\la c|V_{\vecb}^{(e)}|n_2\ra
\la n_1 n_2|I(\om)|c v'\ra 
\nonumber \\
&+\la v|W|n_1\ra\la n_1 c|I(\om)|n_2 v'\ra 
\la n_2|V_{\vecb}^{(e)}|c\ra
\biggr]
\nonumber \\
&+\sum_{n_1,n_2}^{(\veps_{n_1}\neq \veps_c,\,\veps_{n_2}\neq \veps_c)}
\frac{2}{(\veps_c-\veps_{n_1})(\veps_c-\veps_{n_2})}
\biggl[
\la c|V_{\vecb}^{(e)}|n_1\ra\la n_1|W|n_2\ra
\la n_2 v|I(\om)|v' c\ra 
\nonumber \\
&+\la c|V_{\vecb}^{(e)}|n_1\ra\la n_1 v|I(\om)|v' n_2\ra
\la n_2|W|c\ra
+\la c|W|n_1\ra\la n_1|V_{\vecb}^{(e)}|n_2\ra
\la n_2 v|I(\om)|v' c\ra 
\biggr]
\nonumber \\
&-\sum_{\veps_{\tc}=E_{1,-1}}\sum_{n}^{(\veps_{n}\neq \veps_c)}
\frac{2}{(\veps_c-\veps_{n})^2}
\biggl[
\la c|V_{\vecb}^{(e)}|n \ra\la n|W|\tc\ra
\la \tc v|I(\om)|v' c\ra 
\nonumber \\
&+\la c|V_{\vecb}^{(e)}|n \ra
\la n v|I(\om)|v' \tc\ra \la \tc|W|c\ra
+\la c|V_{\vecb}^{(e)}|\tc \ra\la \tc|W|n\ra
\la n v|I(\om)|v' c\ra
\biggr]
\nonumber \\
&-\sum_{\veps_{\tc}=E_{1,-1}}\sum_{n}^{(\veps_{n}\neq \veps_c)}
\frac{2}{\veps_c-\veps_{n}}
\biggl[
\la c|V_{\vecb}^{(e)}|n\ra\la n|W|\tc\ra
\la \tc v|I'(\om)|v' c\ra
\nonumber \\
&+\la c|V_{\vecb}^{(e)}|n\ra
\la n v|I'(\om)|v' \tc\ra\la \tc|W|c\ra
+\la c|V_{\vecb}^{(e)}|\tc\ra\la \tc|W|n\ra
\la n v|I'(\om)|v' c\ra
\biggr]
\nonumber \\
&-\sum_{\veps_{\tc}=E_{1,-1}}\sum_{n}^{(\veps_{n}\neq \veps_v)}
\frac{2}{\veps_v-\veps_{n}}
\biggl[
\la v|V_{\vecb}^{(e)}|n\ra\la n c|I'(\om)|\tc v'\ra
\la \tc|W|c\ra
+\la v|W|n\ra\la n c|I'(\om)|\tc v'\ra
\la \tc|V_{\vecb}^{(e)}|c\ra
\biggr]
\nonumber \\
&+\sum_{\veps_{\tv}=\veps_v}\sum_{n}^{(\veps_{n}\neq \veps_c)}
\frac{2}{\veps_c-\veps_{n}}
\biggl[
\la v|V_{\vecb}^{(e)}|\tv\ra\la \tv c|I'(\om)|n v'\ra
\la n|W|c\ra
+\la v|W|\tv\ra\la \tv c|I'(\om)|n v'\ra
\la n|V_{\vecb}^{(e)}|c\ra
\biggr]
\nonumber \\
&-\sum_{\veps_{\tv}=\veps_v}\sum_{\veps_{\tc}=E_{1,-1}}
\biggl[
\la v|V_{\vecb}^{(e)}|\tv\ra\la \tv c|I''(\om)|\tc v'\ra
\la \tc|W|c\ra
+\la v|W|\tv\ra\la \tv c|I''(\om)|\tc v'\ra
\la \tc|V_{\vecb}^{(e)}|c\ra
\biggr]
\nonumber \\
&+\sum_{\veps_{\tc}=E_{1,-1}}\sum_{\veps_{\bc}=E_{1,-1}}
\la c|V_{\vecb}^{(e)}|\tc\ra\la \tc v|I''(\om)|v' \bc\ra
\la \bc|W|c\ra
 \biggr)\,.
\end{align}
Here 
\begin{eqnarray} \label{f}
f=
\begin{cases}
        1            &\text{for $W=V_{\rm HFS}^{(\mu)}$ 
or $W=V_{\rm HFS}^{(Q)}$}\\
  \frac{1}{2}        &\text{for $W=V_{\vecb}^{(e)}$}
\end{cases}\,,
\end{eqnarray}
\begin{equation}
\la n_1 n_2|I(\om)|n_3 n_4 \ra \equiv \int
d{\vec{x}}_1 d{\vec{x}}_2 u_{n_1}^{\dag}({\vec{x}}_1)
u_{n_2}^{\dag}({\vec{x}}_2)
I(\om) u_{n_3}({\vec{x}}_1)u_{n_4}({\vec{x}}_2)\,,
\end{equation}
\begin{equation}
I(\om)=\alpha\frac{(1-\vec{\alpha}_1\cdot\vec{\alpha}_2)
\cos(\omega r_{12})}{r_{12}}\,,
\end{equation}
\begin{equation}
I'(\om)=\frac{dI(\om)}{d\om}\,,
\,I''(\om)=\frac{d^2 I(\om)}{d{\om}^2}\,,
\end{equation}
$\omega=\veps_v-E_{1,-1}$, and $r_{12}=|{\vec{x}}_1-{\vec{x}}_2|$.
As in the case of evaluation of $\delta g_{\rm HFS}^{(2s)}$ 
considered above, the diagrams corresponding to Eq. (\ref{DeE2CC}) 
are presented in Fig. 1 and the ones corresponding to 
Eq. (\ref{DeE3CC}) are presented in Figs. 3 -- 5. 
Separating the interelectronic-interaction 
corrections is carried out according to Fig. \ref{E-plane} 
just as it was done in section \ref{HFScorr}.
  
The calculation of $h_1(k)$ was discussed in detail in Ref.
\cite{mosk07os}. We found that 
\begin{align}
h_1(k)B&=\mf
\biggl[g_j Y_{\rm el}(k)-\git Y_{\rm nuc}^{(\mu)}(k) \notag  \\
&+\alpha^2 Z \frac{1}{12}
\biggl\{ \git Y_{\rm nuc}^{(\mu)}(k) S_2^{(\rm t)}(\az)+
(\az)^2\frac{23}{180}Q\left(\frac{\me c}{\hbar}\right)^2
Y_{\rm nuc}^{(Q)}(k) T_2^{(\rm t)}(\az)\biggr\}\biggr]
\mub B\,.
\end{align}
Calculating the other matrix elements, we obtain
\begin{align}
h_2(k)B^2&=\frac{14}{(\az)^2}U_2^{(\rm t)}(\az)(\mub B)^2/(\me c^2) 
\,,  \\
\tilde{h}_1(j,k)B&= \frac{1}{2}\frac{\sqrt{(I+1/2)^2-\mf^2}}{I+1/2}
\biggl[g_j+\git \notag  \\
&-\alpha^2 Z \frac{1}{12}\biggl\{ \git S_2^{(\rm t)}(\az)+
(\az)^2\frac{23}{360}Q\left(\frac{\me c}{\hbar}\right)^2
\frac{2I+3}{2I}T_2^{(\rm t)}(\az)\biggr\}\biggr]\mub B\,,   \\
\tilde{h}_2(j,k)&=0 \,.
\end{align}
Here the total functions are
\begin{align}
\label{Utot} 
U_2^{(\rm t)}(\az)&=U_2(\az) + \frac{1}{Z} B_{\vecb}(\az) + 
\frac{1}{Z^2} C_{\vecb}(\az)+\dots\,, \\
\label{Stot}
S_2^{(\rm t)}(\az)&=S_2(\az) + \frac{1}{Z} B_{\mu}(\az) + 
\frac{1}{Z^2} C_{\mu}(\az)+\dots\,,   \\
\label{Ttot}
T_2^{(\rm t)}(\az)&=T_2(\az) + \frac{1}{Z} B_{Q}(\az) + 
\frac{1}{Z^2} C_{Q}(\az)+\dots\,.      
\end{align}
It must be stressed that the contributions of order $1/Z^2$
and higher to these functions are not included in 
Eq. (\ref{secur}). Their evaluation would require 
consideration of some higher-order terms
in operator (\ref{H}). The expansions (\ref{Utot}) -- (\ref{Ttot})
are presented by anology with Eqs. (\ref{gcorrdip}) and (\ref{gcorrquadr}).
The function
\begin{equation}\label{defU2}
U_2(\az)=\frac{(\az)^2 \me c^2}{14(\mub B)^2}\,
\De E_{(\vecb)}^{(2)}(F,F)
\end{equation}
determines the one-electron contribution.
Calculations employing Eqs. (\ref{Ia})~-- (\ref{Id})
yield for the interelectronic-interaction corrections $B_{\vecb}(\az)$,
$B_{\mu}(\az)$, and $B_{Q}(\az)$ 
\begin{align}
\label{defBB}
B_{\vecb}(\az)&=\frac{\alpha^2 Z^3 \me c^2}{14(\mub B)^2}\,
\De E_{(\vecb)}^{(3)}(F,F)\,,                \\
\label{defBmu}
B_{\mu}(\az)&=\frac{12}{\alpha^2\,\frac{\me}{\mpr}g_I
Y_{\rm nuc}^{(\mu)}(F)}\,
\frac{\De E_{(\mu)}^{(3)}(F,F)}{\mub B\mf}\,, \\
\label{defBQ}
B_{Q}(\az)&=\frac{2160}{23 \alpha^4 Z^2\, Q
\left(\frac{\me c}{\hbar}\right)^2
Y_{\rm nuc}^{(Q)}(F)}\,
\frac{\De E_{(Q)}^{(3)}(F,F)}{\mub B\mf}\,.    
\end{align}
It must be noted that, because of the smallness of the
contribution determined by $B_{Q}(\az)$, 
only $B_{\vecb}(\az)$ and $B_{\mu}(\az)$
were evaluated in the present paper.
For checking purposes the calculation of these 
functions was performed in both Feynman and Coulomb gauges.
The results of both calculations coincide with each other.

Solving equation (\ref{secur}), we finally obtain for $\mf=-I+1/2,
...,I-1/2$,
\begin{align}\label{BR3}
\De E_{\rm mag}(x) &= \dhfs^{(2s)}\biggl[a_1(1+\eps_1)\mf x +
 \eps_2 \frac{\dhfs^{(2s)}}{\me c^2}x^2 \notag \\
       & \pm
\frac{1}{2} \sqrt{1 +\frac{4\mf}{2I+1}c_1(1+\delta_1) x
+c_2(1+\delta_2+\mf^2\delta_3) x^2}\,\,\biggr]\,.
\end{align}
Here
\begin{align}
\eps_1&=-\frac{1}{2\git}[\delta
g_{\rm HFS}^{(2s)}(F)+ \delta g_{\rm HFS}^{(2s)}(F+1)] \notag \\
       &= -\alpha^2 Z \frac{1}{12}\left[S_2^{(\rm t)}(\az)-
 (\az)^2\frac{23Q}{120\git}\left(\frac{\me c}{\hbar}\right)^2
\frac{1}{I(2I-1)}T_2^{(\rm t)}(\az)\right] \,,  \\
\eps_2&=\frac{14}{(\az)^2}U_2^{(\rm t)}(\az)\,,
\\ \delta_1&=\frac{2I+1}{2(g_j+\git)}[\delta g_{\rm HFS}^{(2s)}(F+1)
- \delta g_{\rm HFS}^{(2s)}(F)] \notag \\
 &=-\alpha^2 Z \frac{1}{12(g_j+\git)}\left[\git S_2^{(\rm t)}(\az)-
(\az)^2\frac{23}{360}Q\left(\frac{\me c}{\hbar}\right)^2
\frac{4I^2+4I+3}{I(2I-1)}T_2^{(\rm t)}(\az)\right] \,,  \\
\delta_2&= -\alpha^2 Z \frac{1}{6(g_j+\git)}\left[\git 
S_2^{(\rm t)}(\az)+
(\az)^2\frac{23}{360}Q\left(\frac{\me c}{\hbar}\right)^2
\frac{2I+3}{2I}T_2^{(\rm t)}(\az)\right] \,,  \\
\delta_3&= \frac{1}{g_j+\git}\alpha^4 Z^3
\frac{23}{360}Q\left(\frac{\me c}{\hbar}\right)^2
\frac{1}{I(2I-1)}T_2^{(\rm t)}(\az) \,.
\end{align}

For $\Ft=I+\frac{1}{2}$ and $\mf=\pm(I+\frac{1}{2})$, in contrast to
Eq. (\ref{BRl1}), we have
\begin{equation}\label{BRnl}
\De E_{\rm mag}(x)=\dhfs^{(2s)}\biggl[\frac{1}{2} \pm
d_1(1+\eta_1) x+ \eta _2\frac{\dhfs^{(2s)}}{\me c^2} x^2\biggr]\,,
\end{equation}
where
\begin{align}
\eta_1 &=  \alpha^2 Z \frac{1}{6(g_j-2I\git)}\left[\git I 
S_2^{(\rm t)}(\az)+
(\az)^2\frac{23}{360}Q\left(\frac{\me c}{\hbar}\right)^2
T_2^{(\rm t)}(\az)\right] \,, \\ 
\eta _2 &=\eps_2=\frac{14}{(\az)^2} 
U_2^{(\rm t)}(\az)\,,
\end{align}
and the ``$-$'' and ``$+$'' signs correspond to
$\mf=-(I+\frac{1}{2})$ and $\mf=I+\frac{1}{2}$, respectively.

If $I=1/2$, the electrical quadrupole interaction vanishes and one
can easily obtain for $\mf=0$:
\begin{equation}\label{BR4}
\De E_{\rm mag}(x)=\dhfs^{(2s)}\biggl[\eps_2
\frac{\dhfs^{(2s)}}{\me c^2}x^2
        \pm
\frac{1}{2} \sqrt{1 +c_2(1+\delta_2)x^2 }\,\biggr]
\end{equation}
with
\begin{equation}
\delta_2= -\frac{\git}{6(g_j+\git)}\alpha^2 Z  S_2^{(\rm t)}(\az)\,.
\end{equation}
For $I=1/2$, $\mf=\pm 1$, the effect is described by formula
(\ref{BRnl}) with
\begin{equation}
\eta_1 = \frac{\git}{12(g_j-\git)}\alpha^2 Z S_2^{(\rm t)}(\az)\,.
\end{equation}

\section{Numerical results}\label{Numer}

In Table 1, we present the numerical results for the functions
$U_2(\az)$, $B_{\vecb}(\az)$, $U_2^{(t)}(\az)$,
$S_2(\az)$, $B_{\mu}(\az)$, $S_2^{(t)}(\az)$,  and $T_2(\az)$ (only
for the isotopes with $I>1/2$) defined by Eqs. (\ref{defU2}),
(\ref{defBB}), (\ref{Utot}), (\ref{defS2}), (\ref{defBmu}), (\ref{Stot}),
and (\ref{defT2}), respectively, for the $2s$ state. 
All the values are calculated 
for the extended nuclear charge distribution. The root-mean-square 
nuclear charge radii $\la r^2\ra^{1/2}$ were taken from 
Ref. \cite{RnewAng}. For those elements for which no accurate
experimental radii were available we employed the empirical
expression \cite{joh85}
\begin{align}
\langle r^2\rangle^{1/2}=0.836 A^{1/3}+0.570(\pm 0.05)\
\text{fm}\,,
\end{align}
where $A$ is the nuclear mass expressed in a.m.u.  
The calculations were performed using the
dual-kinetic-balance (DKB) basis set method \cite{DKB04}
with the basis functions constructed from B-splines \cite{joh86,
joh88}. The uncertainties of $U_2(\az)$, $B_{\vecb}(\az)$, 
$S_2(\az)$, $B_{\mu}(\az)$, and $T_2(\az)$ were estimated by 
adding quadratically two errors, one obtained by varying 
$\la r^2\ra^{1/2}$ within its 
uncertainty and the other obtained by changing the model of the 
nuclear-charge distribution from the Fermi to the 
homogeneously-charged-sphere model.
The uncertainties of the total functions $U_2^{(t)}(\az)$ 
and $S_2^{(t)}(\az)$ due to uncalculated second- and higher-order 
terms were estimated as the first-order correction 
($\sim B_{\vecb}(\az)/Z$ and $\sim B_{\mu}(\az)/Z$, respectively) 
multiplied by the factor $2/Z$. The
uncertainty due to uncalculated first- and higher-order terms 
in Eq. (\ref{Ttot}) was estimated in a similar way.

In Table 2, we present the individual contributions to the $2s$
$g_j$ factor for some Li-like ions with $I\neq 0$ in the range
$Z=6-32$.  The Dirac point-nucleus value is obtained by
Eq. (\ref{gD}). The interelectronic-interaction ($\De g_{\rm int}$), 
QED ($\Delta g_{\rm QED}$), nuclear-recoil ($\Delta g_{\rm rec}^{(e)}$), 
and nuclear-size ($\Delta g_{\rm NS}$) corrections are obtained 
as described in Refs. \cite{glasha04,glavol06}. The
nuclear-size correction was evaluated for the
homogeneously-charged-sphere model if $Z=6-16$ and
for the Fermi model if $Z=20-32$ . The nuclear polarization
contribution to the ${2s}$ $g_j$ factor of light Li-like ions
can be neglected \cite{nef02}. The $g_j$ factor values
given in Table 2 are used for calculations of the coefficients 
in the Breit -- Rabi formula.

In Table 3, the numerical results for the coefficients in Eqs.
(\ref{BR1}), (\ref{BRl1}), (\ref{BR2}), (\ref{BR3}), 
(\ref{BRnl}), and (\ref{BR4}) are
listed  for some Li-like isotopes in the interval
$Z=6-32$. Since in all the cases under consideration the absolute 
value of the recoil correction to the bound-nucleus $g_I$ factor 
is smaller than $10^{-11}$ \cite{mar01}, we actually have 
in Eq. (\ref{defgit}): $\git=\frac{\me}{\mpr}g_I$.

\section{Discussion}

The energy separation between the ground-state HFS components
($F=I-1/2$ and $F^{\prime}=I+1/2$) of a lithiumlike ion can be
written as \cite{sha95}
\begin{align}\label{ShHFS}
\dhfs^{(2s)}&=\frac{1}{6}\alpha (\az)^3\frac{\mu}{\mun}\frac{\me}{\mpr}
\frac{2I+1}{2I}\me c^2 \notag  \\
&\times \biggl\{[A^{(2s)}(\az)(1-\delta^{(2s)}
(1-\epsilon^{(2s)})+x_{\rm rad}^{(2s)}]  \notag \\
&+ \frac{1}{Z} B^{(2s)}(\az) + 
\frac{1}{Z^2} C^{(2s)}(\az)+\dots\biggr\}  \,,
\end{align}
where
\begin{equation}\label{A(2s)}
A^{(2s)}(\az)=\frac{2[2(1+\ga)+\sqrt{2(1+\ga)}]}
{(1+\ga)^2\ga(4\ga^2-1)}=1+\frac{17}{8}(\az)^2
+\frac{449}{128}(\az)^4+\dots
\end{equation}
is the one-electron relativistic factor, $\delta^{(2s)}$ 
is the nuclear charge
distribution correction, $\epsilon^{(2s)}$ is the nuclear
magnetization distribution correction (the Bohr -- Weisskopf
effect), $x_{\rm rad}^{(2s)}$ is the QED correction,
$B^{(2s)}(\az)$ and $C^{(2s)}(\az)$ determine the
interelectronic-interaction corrections to the hyperfine structure.
Therefore, the dimensionless variable $x=\mub B/\dhfs^{(2s)}$ is
of order of $6\mub B/[\alpha (\az)^3\frac{\me}{\mpr}\me
c^2]$. The intervals of $B$ and $Z$, for which $x\sim 1$, are of special
interest (in the original paper \cite{BR31} the fields with
$0.1\leqslant x\leqslant 3$ were considered to be intermediate).

For the magnetic fields with the magnitude $B\sim 1-10 \, T$, that
are generally used in this kind of experiments, Li-like ions with
$Z=6-32$ meet the requirement $x\sim 1$. For this reason, only
such ions are presented in Tables 1 -- 3.

For ions with $Z\leq 32$, the electric-quadrupole corrections
to the coefficients $a_1$, $c_1$, $c_2$, and $d_1$ are either 
equal to zero, if $I=1/2$, as in the case of $^{13}\rm C^{3+}$,
or by a factor of $10^{-3}-10^{-4}$ smaller than the magnetic-dipole ones.
This is due to an additional factor $(\alpha Z)^2$ in the 
electric-quadrupole contributions compared to the magnetic-dipole ones 
in the equations for the hyperfine-structure corrections to the
Breit -- Rabi formula coefficients and small
values of $Q$ for low-$Z$ ions.

As one can see from Table 3, the corrections $\epsilon_1$,
$\delta_1$, $\delta_2$, $\delta_3$, and $\eta_1$ for Li-like ions 
are several times smaller as compared to the corresponding ones 
for the $1s$ state of the same H-like isotopes \cite{mosk06}. 
However, they provide  more precise
determinations of the coefficients in the Breit -- Rabi formula.

For $B= 1-10 \, T$, an estimate of the terms of the third 
and higher orders with respect to $B$ in Eq. (\ref{secur}) 
indicates that the
contributions from these terms are negligibly small as compared
to both magnetic dipole and electrical quadrupole corrections.
However, it is very important to take into account $\epsilon_2
B^2$ and $\eta_2 B^2$ if $Z=6-32$. This is due to the fact that
these terms are comparable with the other
corrections to the Breit -- Rabi formula considered and the less
$Z$ is, the more appreciable the contributions from $\epsilon_2
B^2$ and $\eta_2 B^2$ become. One can see that for Li-like ions 
these terms are $10-10^3$ times bigger as compared to 
the case of the $1s$ state of the same H-like isotopes \cite{mosk06}.
In the second-order approximation (\ref{secur}) with respest to $B$, 
formulas (\ref{BR1}), (\ref{BR2}), (\ref{BR3}), and (\ref{BR4}) do not 
contain $B$ to a power higher than two under the square root.
This is due to the fact that $h_2(F)=h_2(\Ft)$. 

The Breit -- Rabi formula for the $2s$ state contains
$\dhfs^{(2s)}$, and the coefficients in the formula and
the corrections to them calculated above include the value of
$\mu/\mun$. The uncertainties of the nuclear magnetic moments 
indicated in Table 3, as a rule, do not include errors due to 
unknown chemical shifts which, in some cases, can contribute 
on the level of a few tenths percents. Thus, carrying out 
the experiments on the Zeeman splitting with the aforesaid 
accuracy could provide the most precise determination of both 
$\dhfs^{(2s)}$ and $\mu/\mun$.
The corrections to the Breit -- Rabi formula evaluated
in this paper will be important for this determination.

\section{Acknowledgements}
D.L.M. thanks the support by INTAS-GSI grant No. 05-111-4937.
He is also grateful to GSI.
V.M.S. acknowledges the support by INTAS-GSI grant No. 06-1000012-8881. 
This work was also supported by RFBR (grant No. 07-02-00126) and DFG.

\clearpage
\newpage

\begin{table}
\caption{The numerical results for the extended-charge-nucleus 
values of functions $U_2^{(\rm t)}(\az)$, $S_2^{(\rm t)}(\az)$, 
and $T_2^{(\rm t)}(\az)$ (for the ions with $I\neq 1/2$). 
The values of $\la r^2\ra^{1/2}$ are taken from Ref. \cite{RnewAng}.}
\begin{center}
\begin{tabular}{||c|l|l|l|l|l|l|l||}
\hline
Ion & $^{13}\rm{C}^{3+}$ & $^{17}\rm{O}^{5+}$
& $^{21}\rm{Ne}^{7+}$ & $^{25}\rm{Mg}^{9+}$ 
& $^{33}\rm{S}^{13+}$ & $^{43}\rm{Ca}^{17+}$ 
& $^{53}\rm{Cr}^{21+}$   
\\ \hline
$Z$ &6 &8 &10 &12 &16 &20 &24     
\\ \hline
$\la r^2\ra^{1/2}$, fm &2.461 &2.695 &2.967 &3.028 
                      &3.251 &3.493 &3.659  
\\\hline 
$U_2(\az)$ &0.998574 &0.997464 &0.996038  &0.994295
           &0.989858 &0.984153 &0.977179  
\\ \hline
$B_{\vecb}(\az)$   &2.47400 &2.47359 &2.47305  &2.47240
                  &2.47070 &2.46848 &2.46568  
\\ \hline
$U_2^{(\rm t)}(\az)$        &1.41(14)  &1.31(8)   &1.24(5)  &1.20(3) 
                           &1.144(19) &1.108(12) &1.080(9)  
\\ \hline
$S_2(\az)$ &1.00306 &1.00545 &1.00854  &1.01235
           &1.02218 &1.03513 &1.05145  
\\ \hline
$B_{\mu}(\az)$    &-1.60040    &-1.60757    &-1.61684  &-1.62825     
                 &-1.65769(1) &-1.69639(1) &-1.74505(2)  
\\ \hline
$S_2^{(\rm t)}(\az)$  &0.74(9)  &0.80(5)  &0.85(3)  &0.88(2)     
                     &0.919(13)  &0.950(8) &0.979(6)  
\\ \hline
$T_2(\az)$ &------------ &1.00448(2) &1.00710(2)  &1.01051(3)
           & 1.01927(4)  &1.03083(7) &1.04531(10) 
\\ \hline
$T_2^{(\rm t)}(\az)$ &------------ &1.0(3)   &1.0(2)  &1.01(17)
                    & 1.02(13)    &1.03(10) &1.05(9)     
\\ \hline
\end{tabular}
\end{center}
\begin{center}
\begin{tabular}{||c|l|l|l||}
\hline
Ion                   &$^{61}\rm{Ni}^{25+}$ 
&$^{67}\rm{Zn}^{27+}$  &$^{73}\rm{Ge}^{29+}$ 
\\ \hline
$Z$  &28 &30 &32    
\\ \hline
$\la r^2\ra^{1/2}$, fm  &3.822 &3.964 &4.063 
\\\hline 
$U_2(\az)$ &0.968938 &0.964342 &0.959428 
\\ \hline
$B_{\vecb}(\az)$   &2.46226 &2.46030 &2.45817 
\\ \hline
$U_2^{(\rm t)}(\az)$    &1.057(6) &1.046(5) &1.036(5) 
\\ \hline
$S_2(\az)$    &1.07146(1) & 1.08296(1) &1.09555(2) 
\\ \hline
$B_{\mu}(\az)$   &-1.80458(3) &-1.83874(3)  &-1.87609(4) 
\\ \hline
$S_2^{(\rm t)}(\az)$  &1.007(5) &1.022(4) &1.037(4) 
\\ \hline
$T_2(\az)$     &1.06287(11) &1.07277(14)    &1.08359(13) 
\\ \hline
$T_2^{(\rm t)}(\az)$   &1.06(8) &1.07(7) &1.08(7) 
\\ \hline
\end{tabular}
\end{center}
\end{table}

\begin{table}
\caption{  The individual contributions to the ground-state 
$g_j$ factor of lithiumlike ions with nonzero nuclear spin 
and the nuclear charge in the range $Z=6-32$.
The values of $\la r^2\ra^{1/2}$ are the same as in Table 1. }
\begin{center}
\begin{tabular}{||c|r@{.}l|r@{.}l|r@{.}l|r@{.}l||}
\hline
 Ion
&\multicolumn{2}{c|}{$^{13}\rm{C}^{3+}$}
&\multicolumn{2}{c|}{$^{17}\rm{O}^{5+}$}
&\multicolumn{2}{c|}{ $^{33}\rm{S}^{13+}$}
&\multicolumn{2}{c||}{$^{43}\rm{Ca}^{17+}$} 
\\ \hline
 $g_{\rm D}$
 &1&999680300  &1&999431380      
 &1&997718193  &1&996426011
\\ \hline

 $\De g_{\rm{int}}$
 &0&000130758(19)      &0&00017666(3)
 &0&00036124(9)        &0&00045445(14)
\\ \hline

$\De g_{\rm{QED}}$
 &0&002319417(6)  &0&002319549(12)      
 &0&00232070(6)   &0&00232171(10)
 \\ \hline

$\De g_{\rm{rec}}^{(e)}$
 &0&000000009      &0&000000016
 &0&000000045(1)   &0&000000057(2) 
\\ \hline

 $\De g_{\rm{NS}}$
 &0&0    &0&0     
 &0&000000005   &0&000000014
\\ \hline

 $g_j$
&2&00213048(2)    &2&00192760(3)
&2&00040018(11)   &1&99920224(17) \\ \hline
\end{tabular}
\end{center}
\begin{center}
\begin{tabular}{||c|r@{.}l|r@{.}l||}
\hline
 Ion
&\multicolumn{2}{c|}{$^{53}\rm{Cr}^{21+}$}
&\multicolumn{2}{c||}{$^{73}\rm{Ge}^{29+}$}
\\ \hline

 $g_{\rm D}$
&1&994838064 &1&990752307
\\ \hline

 $\De g_{\rm{int}}$
&0&0005485(2) &0&0007397(4)
\\ \hline

$\De g_{\rm{QED}}$
&0&00232304(15)  &0&0023270(2)
 \\ \hline

$\De g_{\rm{rec}}^{(e)}$
&0&000000069(4)  &0&000000093(9) 
\\ \hline

 $\De g_{\rm{NS}}$
&0&000000035  &0&000000160
\\ \hline

 $g_j$
&1&9977097(3)   &1&9938193(4) \\ \hline
\end{tabular}
\end{center}
\end{table}

\begin{table}
\caption{The numerical values of the coefficients in Eqs. (\ref{BR1}),
(\ref{BRl1}), (\ref{BR2}), (\ref{BR3}), (\ref{BRnl}),
and (\ref{BR4})
for Li-like ions with $Z=6-32$.
The values of $\mu/\mun$ and $Q$ are taken from Refs. \cite{rag89}
and \cite{PPyy01}, respectively.}
\begin{center}
\begin{tabular}{||c|r@{.}l|r@{.}l|r@{.}l|r@{.}l||}
\hline
 Ion
&\multicolumn{2}{c|}{$^{13}\rm{C}^{3+}$}
&\multicolumn{2}{c|}{$^{17}\rm{O}^{5+}$}
&\multicolumn{2}{c|}{ $^{33}\rm{S}^{13+}$}
&\multicolumn{2}{c||}{$^{43}\rm{Ca}^{17+}$} 
\\ \hline
$I$
&\multicolumn{2}{c|}{1/2} 
&\multicolumn{2}{c|}{5/2}
&\multicolumn{2}{c|}{3/2}
&\multicolumn{2}{c||}{7/2}
\\ \hline
$\mu/\mun$
&\multicolumn{2}{c|}{0.7024118(14)} 
&\multicolumn{2}{c|}{-1.89379(9)}
&\multicolumn{2}{c|}{0.6438212(14)}
&\multicolumn{2}{c||}{-1.317643(7)} 
\\ \hline
$Q$, barn
&\multicolumn{2}{c|}{------}
&\multicolumn{2}{c|}{-0.02558(22)}
&\multicolumn{2}{c|}{-0.0678(13)}
&\multicolumn{2}{c||}{-0.0408(8)} 
\\ \hline
$a_1$ 
&\multicolumn{2}{c|}{------}
&0&00041256(2)   
&-0&0002337573(5) 
&0&0002050317(11) 
\\ \hline
$\epsilon_1$
&\multicolumn{2}{c|}{------}
&-0&0000284(18)
&-0&0000653(9)
&-0&0000843(7)
\\ \hline
$a_1(1+\epsilon_1)$ 
&\multicolumn{2}{c|}{------}
&0&00041254(2) 
&-0&0002337421(6) 
&0&0002050144(11) 
\\ \hline
$\epsilon_2(=\eta_2)$
&\multicolumn{2}{c|}{1.03(10)$\times 10^4$} 
&\multicolumn{2}{c|}{5.4(3)$\times 10^3$}
&\multicolumn{2}{c|}{1.17(2)$\times 10^3$}
&\multicolumn{2}{c||}{7.28(8)$\times 10^2$} 
\\ \hline
$c_1$ &\multicolumn{2}{c|}{------}
&2&00151505(4) 
&2&00063394(11)  
&1&99899721(17)  
\\ \hline
$\delta_1$ 
&\multicolumn{2}{c|}{------} 
&0&0000000059(4)
&-0&00000000763(11)
&0&00000000864(7)
\\ \hline
$c_1(1+\delta_1)$ 
&\multicolumn{2}{c|}{------}
&2&00151506(4) 
&2&00063393(11) 
&1&99899722(17) 
\\ \hline
$c_2$
&4&01159069(8)  
&4&00606248(15) 
&4&0025362(4) 
&3&9959898(7)  
\\ \hline
$\delta_2$ 
&-0&0000000151(18)
&0&0000000117(7)
&-0&0000000152(2)
&0&00000001730(15)
\\ \hline
$\delta_3$ 
&\multicolumn{2}{c|}{------}
&0&0 
&-0&00000000001
&0&0  
\\ \hline
$c_2(1+\delta_2)$ 
&4&01159062(8)
&4&00606253(15)
&4&0025361(4) 
&3&9959899(7) 
\\ \hline
$c_2\delta_3$ 
&\multicolumn{2}{c|}{------}
&0&0 
&-0&00000000002(1)
&0&0  
\\ \hline
$d_1$
&1&000682697(10) 
&1&00199519(5) 
&0&99984946(5) 
&1&00031873(9) 
\\ \hline
$\eta_1$
&0&0000000075(9)
&-0&0000000292(18) 
&0&0000000229(3)
&-0&0000000605(5) 
\\ \hline
$d_1(1+\eta_1)$ 
&1&000682704(10) 
&1&00199516(5) 
&0&99984948(5)  
&1&00031867(9) 
\\ \hline
\end{tabular}
\end{center}
\begin{center}
\begin{tabular}{||c|r@{.}l|r@{.}l||}
\hline
 Ion
&\multicolumn{2}{c|}{$^{53}\rm{Cr}^{21+}$}
&\multicolumn{2}{c||}{$^{73}\rm{Ge}^{29+}$}
 \\
\hline
$I$
&\multicolumn{2}{c|}{3/2}
&\multicolumn{2}{c||}{9/2}
  \\
\hline
$\mu/\mun$
&\multicolumn{2}{c|}{-0.47454(3)} 
&\multicolumn{2}{c||}{-0.8794677(2)}
\\ \hline
$Q$, barn
&\multicolumn{2}{c|}{-0.150(50)}
&\multicolumn{2}{c||}{-0.196}
\\ \hline
$a_1$   
&0&000172295(11)
&0&00010643846(2)
\\ \hline
$\epsilon_1$
&-0&0001041(6)
&-0&0001472(6)
\\ \hline
$a_1(1+\epsilon_1)$  
&0&000172277(11)
&0&00010642279(7)
\\ \hline
$\epsilon_2(=\eta_2)$
&\multicolumn{2}{c|}{4.93(4)$\times 10^2$}
&\multicolumn{2}{c||}{2.660(13)$\times 10^2$} 
\\ \hline
$c_1$   
&1&9975374(3)
&1&9937129(4)
\\ \hline
$\delta_1$  
&0&00000000893(5)
&0&00000000776(3)
\\ \hline
$c_1(1+\delta_1)$  
&1&9975374(3)
&1&9937129(4)
\\ \hline
$c_2$ 
&3&9901556(10)
&3&9748910(17)
\\ \hline
$\delta_2$  
&0&00000001803(11)
&0&00000001582(6)
\\ \hline
$\delta_3$  
&-0&00000000004(2)
&-0&00000000001
\\ \hline
$c_2(1+\delta_2)$ 
&3&9901557(10)
&3&9748911(17)
\\ \hline
$c_2\delta_3$  
&-0&00000000018(6)
&-0&00000000005
\\ \hline
$d_1$ 
&0&99911329(13)
&0&9973886(2)
\\ \hline
$\eta_1$ 
&-0&00000002699(17)
&-0&0000000708(3)
\\ \hline
$d_1(1+\eta_1)$  
&0&99911326(13)
&0&9973886(2)
\\ \hline
\end{tabular}
\end{center}
\end{table}


\clearpage
\newpage
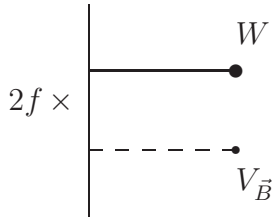
\begin{figure}
\begin{picture}(200,120)
\put(50,0){\line(0,1){80}}
\put(50,55){\line(1,0){55}}
\put(105,55){\circle*{5}}
\multiput(50,25)(10,0){6}{\line(1,0){5}}
\put(105,25){\circle*{3}}
\put(105,65){$W$}
\put(105,10){$V_{\vecb}$}
\put(20,40){$2f\,\times$}
\end{picture}
\caption{The second-order diagrams 
contributing to $S_2^{(\rm t)}(\az)$, 
$T_2^{(\rm t)}(\az)$ (if $f=1$ and 
$W=V_{\rm HFS}^{(\mu)}\,\,\text{or}\,\,W=V_{\rm HFS}^{(Q)}$), 
and $U_2^{(\rm t)}(\az)$ (if $f=\frac{1}{2}$ and 
$W=V_{\vecb}^{(e)}$).}\label{Diagrams2} 
\end{figure}
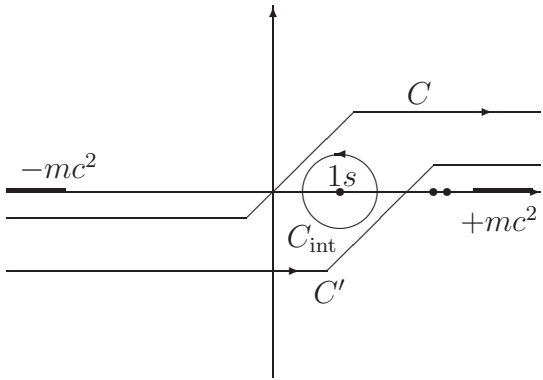
\begin{figure}
\begin{picture}(200,140)
\put(0,70){\vector(1,0){200}}
\put(100,0){\vector(0,1){140}}
\put(125,70){\circle{26}}
\put(125,70){\circle*{3}}
\put(125,84){\vector(-1,0){2}}
\put(105,49){$C_{\rm int}$}
\put(120,72){$1s$}
\put(0,60){\line(1,0){90}}
\put(0,40){\line(1,0){120}}
\put(108,40){\vector(1,0){2}}
\put(115,28){$C'$}
\put(90,60){\line(1,1){40}}
\put(120,40){\line(1,1){40}}
\put(130,100){\line(1,0){70}}
\put(160,80){\line(1,0){40}}
\put(180,100){\vector(1,0){2}}
\put(150,103){$C$}
\multiput(160,70)(5,0){2}{\circle*{3}}
\put(175,71){\line(1,0){22}}
\put(175,70.5){\line(1,0){22}}
\put(170,55){$+mc^2$}
\put(0,71){\line(1,0){22}}
\put(0,70.5){\line(1,0){22}}
\put(5,75){$-mc^2$}
\end{picture}
\caption{$C$ is the original contour of the integration
over the electron energy variable in the formalism with
the standard vacuum. $C'$ is the integration contour for 
the vacuum with the $(1s)^2$ shell included.
The integral along the contour $C_{\rm int}=C'-C$ describes 
the interaction of the valent electron with the $(1s)^2$-shell
electrons.}\label{E-plane} 
\end{figure}
\begin{figure}
\begin{picture}(460,460)
\multiput(50,360)(130,0){3}{\line(0,1){100}}
\multiput(55,430)(20,0){2}{\oval(10,10)[t]}
\multiput(65,430)(20,0){2}{\oval(10,10)[b]}
\put(104,430){\circle{29}}
\put(50,400){\line(1,0){55}}
\put(105,400){\circle*{5}}
\multiput(50,380)(10,0){6}{\line(1,0){5}}
\put(105,380){\circle*{3}}
\put(180,440){\line(1,0){55}}
\put(235,440){\circle*{5}}
\multiput(185,410)(20,0){2}{\oval(10,10)[t]}
\multiput(195,410)(20,0){2}{\oval(10,10)[b]}
\put(234,410){\circle{29}}
\multiput(180,380)(10,0){6}{\line(1,0){5}}
\put(235,380){\circle*{3}}
\multiput(315,430)(20,0){2}{\oval(10,10)[t]}
\multiput(325,430)(20,0){2}{\oval(10,10)[b]}
\put(364,430){\circle{29}}
\multiput(310,400)(10,0){6}{\line(1,0){5}}
\put(365,400){\circle*{3}}
\put(310,380){\line(1,0){55}}
\put(365,380){\circle*{5}}
\put(0,410){$\biggl($}
\multiput(130,410)(130,0){2}{+}
\put(400,410){$\biggr)$}
\put(410,410){$\times\, 2f$}
\put(115,390){$W$}
\put(115,370){$V_{\vecb}$}
\put(460,410){(a)}
\multiput(50,240)(130,0){3}{\line(0,1){100}}
\put(50,330){\oval(10,10)[tr]}
\put(50,290){\oval(10,10)[br]}
\put(60,330){\oval(10,10)[bl]}
\put(60,290){\oval(10,10)[tl]}
\put(60,320){\oval(10,10)[tr]}
\put(60,300){\oval(10,10)[br]}
\put(70,320){\oval(10,10)[bl]}
\put(70,300){\oval(10,10)[tl]}
\put(70,310){\oval(10,10)[tr]}
\put(70,310){\oval(10,10)[br]}
\put(50,270){\line(1,0){55}}
\put(105,270){\circle*{5}}
\multiput(50,250)(10,0){6}{\line(1,0){5}}
\put(105,250){\circle*{3}}
\put(180,310){\oval(10,10)[tr]}
\put(180,270){\oval(10,10)[br]}
\put(190,310){\oval(10,10)[bl]}
\put(190,270){\oval(10,10)[tl]}
\put(190,300){\oval(10,10)[tr]}
\put(190,280){\oval(10,10)[br]}
\put(200,300){\oval(10,10)[bl]}
\put(200,280){\oval(10,10)[tl]}
\put(200,290){\oval(10,10)[tr]}
\put(200,290){\oval(10,10)[br]}
\put(180,330){\line(1,0){55}}
\put(235,330){\circle*{5}}
\multiput(180,250)(10,0){6}{\line(1,0){5}}
\put(235,250){\circle*{3}}
\put(310,330){\oval(10,10)[tr]}
\put(310,290){\oval(10,10)[br]}
\put(320,330){\oval(10,10)[bl]}
\put(320,290){\oval(10,10)[tl]}
\put(320,320){\oval(10,10)[tr]}
\put(320,300){\oval(10,10)[br]}
\put(330,320){\oval(10,10)[bl]}
\put(330,300){\oval(10,10)[tl]}
\put(330,310){\oval(10,10)[tr]}
\put(330,310){\oval(10,10)[br]}
\multiput(310,270)(10,0){6}{\line(1,0){5}}
\put(365,270){\circle*{3}}
\put(310,250){\line(1,0){55}}
\put(365,250){\circle*{5}}
\put(0,290){$\biggl($}
\multiput(130,290)(130,0){2}{+}
\put(400,290){$\biggr)$}
\put(410,290){$\times\, 2f$}
\put(460,290){(b)}
\multiput(50,120)(130,0){3}{\line(0,1){100}}
\multiput(55,190)(20,0){2}{\oval(10,10)[t]}
\put(65,190){\oval(10,10)[b]}
\put(94,190){\circle{29}}
\put(108,190){\line(1,0){45}}
\put(155,190){\circle*{5}}
\multiput(50,150)(10,0){6}{\line(1,0){5}}
\put(105,150){\circle*{3}}
\multiput(185,190)(20,0){2}{\oval(10,10)[t]}
\put(195,190){\oval(10,10)[b]}
\put(224,190){\circle{29}}
\multiput(238,190)(10,0){5}{\line(1,0){5}}
\put(283,190){\circle*{3}}
\put(180,150){\line(1,0){55}}
\put(235,150){\circle*{5}}
\multiput(315,170)(20,0){2}{\oval(10,10)[t]}
\put(325,170){\oval(10,10)[b]}
\put(354,170){\circle{29}}
\put(362,181){\line(0,1){35}}
\put(362,217){\circle*{5}}
\multiput(362,159)(0,-10){4}{\line(0,-1){5}}
\put(362,123){\circle*{3}}
\put(0,170){$\biggl($}
\multiput(130,170)(130,0){2}{+}
\put(400,170){$\biggr)$}
\put(410,170){$\times\, 2f$}
\put(460,170){(c)}
\multiput(50,0)(130,0){3}{\line(0,1){100}}
\put(50,90){\oval(10,10)[tl]}
\put(50,50){\oval(10,10)[bl]}
\put(40,90){\oval(10,10)[br]}
\put(40,50){\oval(10,10)[tr]}
\put(40,80){\oval(10,10)[tl]}
\put(40,60){\oval(10,10)[bl]}
\put(30,80){\oval(10,10)[br]}
\put(30,60){\oval(10,10)[tr]}
\put(30,70){\oval(10,10)[tl]}
\put(30,70){\oval(10,10)[bl]}
\put(50,70){\line(1,0){55}}
\put(105,70){\circle*{5}}
\multiput(50,30)(10,0){6}{\line(1,0){5}}
\put(105,30){\circle*{3}}
\put(180,90){\oval(10,10)[tl]}
\put(180,50){\oval(10,10)[bl]}
\put(170,90){\oval(10,10)[br]}
\put(170,50){\oval(10,10)[tr]}
\put(170,80){\oval(10,10)[tl]}
\put(170,60){\oval(10,10)[bl]}
\put(160,80){\oval(10,10)[br]}
\put(160,60){\oval(10,10)[tr]}
\put(160,70){\oval(10,10)[tl]}
\put(160,70){\oval(10,10)[bl]}
\multiput(180,70)(10,0){6}{\line(1,0){5}}
\put(235,70){\circle*{3}}
\put(180,30){\line(1,0){55}}
\put(235,30){\circle*{5}}
\put(310,70){\oval(10,10)[tl]}
\put(310,30){\oval(10,10)[bl]}
\put(300,70){\oval(10,10)[br]}
\put(300,30){\oval(10,10)[tr]}
\put(300,60){\oval(10,10)[tl]}
\put(300,40){\oval(10,10)[bl]}
\put(290,60){\oval(10,10)[br]}
\put(290,40){\oval(10,10)[tr]}
\put(290,50){\oval(10,10)[tl]}
\put(290,50){\oval(10,10)[bl]}
\put(310,60){\line(1,0){55}}
\put(365,60){\circle*{5}}
\multiput(310,40)(10,0){6}{\line(1,0){5}}
\put(365,40){\circle*{3}}
\put(0,50){$\biggl($}
\multiput(130,50)(130,0){2}{+}
\put(400,50){$\biggr)$}
\put(410,50){$\times\, 2f$}
\put(460,50){(d)}
\end{picture}
\caption{The third-order diagrams 
contributing to $S_2^{(\rm t)}(\az)$ (if $f=1$ and 
$W=V_{\rm HFS}^{(\mu)}$) and 
$U_2^{(\rm t)}(\az)$ (if $f=\frac{1}{2}$ and 
$W=V_{\vecb}^{(e)}$) being combined
with products of the lower-order 
diagrams presented in Figs. \ref{Diagrams31} and \ref{Diagrams32}.}
\label{Diagrams3} 
\end{figure}
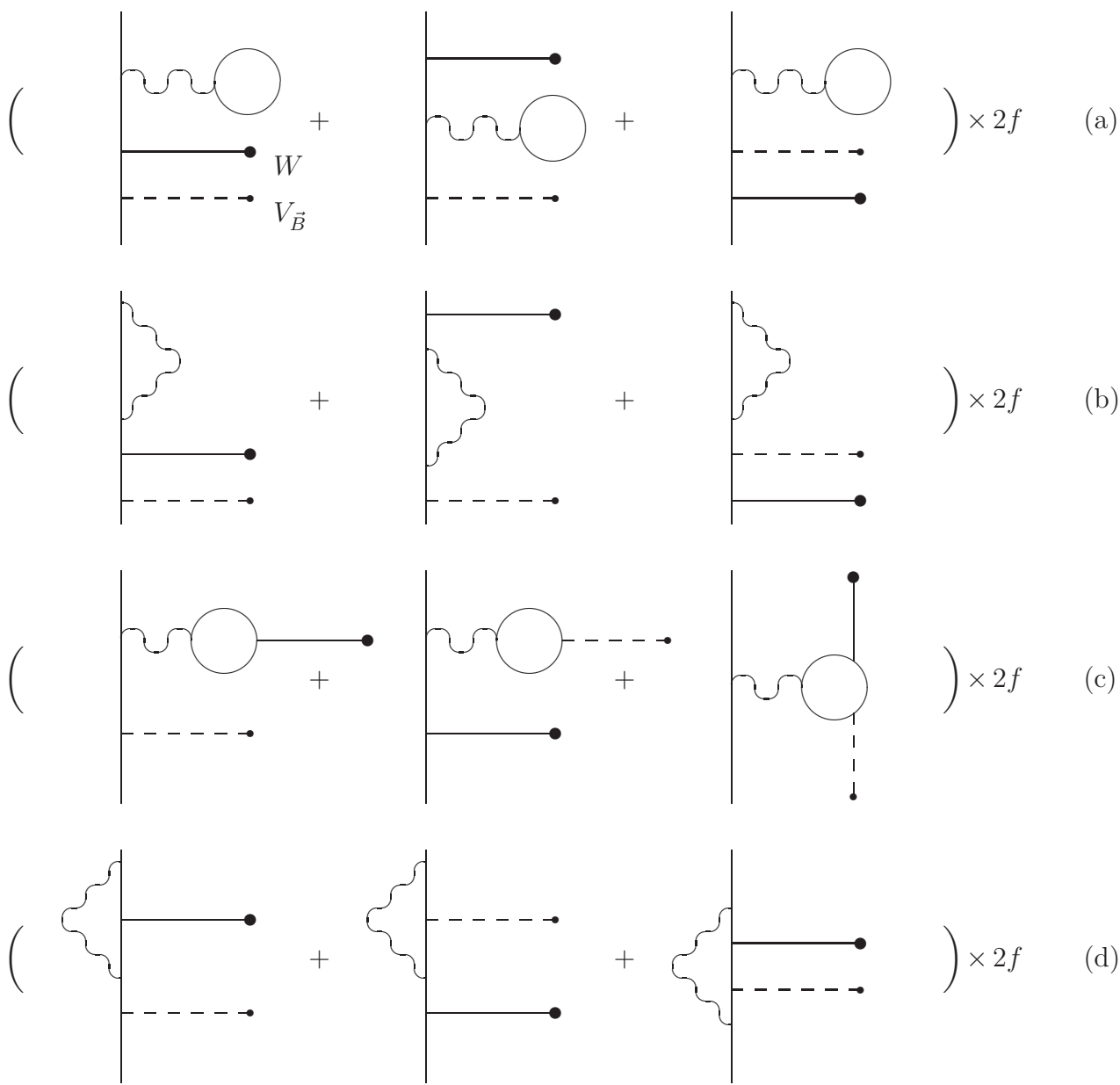
\begin{figure}
\begin{picture}(460,70)
\multiput(50,0)(120,0){4}{\line(0,1){70}}
\put(50,35){\line(1,0){55}}
\put(105,35){\circle*{5}}
\put(105,45){$V_{\rm HFS}^{(\mu)}$}
\multiput(170,35)(10,0){6}{\line(1,0){5}}
\put(225,35){\circle*{3}}
\put(225,45){$V_{\vecb}$}
\multiput(295,35)(20,0){2}{\oval(10,10)[t]}
\multiput(305,35)(20,0){2}{\oval(10,10)[b]}
\put(344,35){\circle{29}}
\put(410,55){\oval(10,10)[tr]}
\put(410,15){\oval(10,10)[br]}
\put(420,55){\oval(10,10)[bl]}
\put(420,15){\oval(10,10)[tl]}
\put(420,45){\oval(10,10)[tr]}
\put(420,25){\oval(10,10)[br]}
\put(430,45){\oval(10,10)[bl]}
\put(430,25){\oval(10,10)[tl]}
\put(430,35){\oval(10,10)[tr]}
\put(430,35){\oval(10,10)[br]}
\end{picture}
\caption{The first-order diagrams 
contributing to $S_2^{(\rm t)}(\az)$ and 
$U_2^{(\rm t)}(\az)$ being multiplied
by the second-order diagrams presented 
in Fig. \ref{Diagrams32}.}\label{Diagrams31} 
\end{figure}
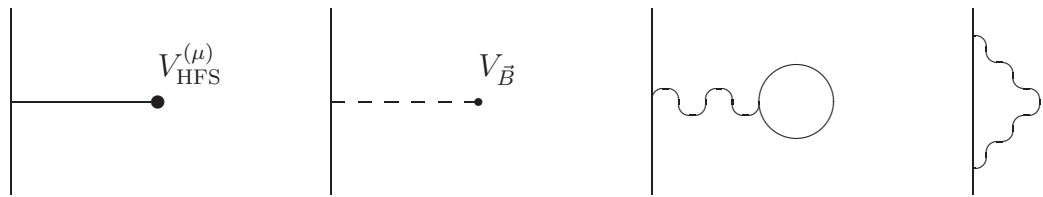
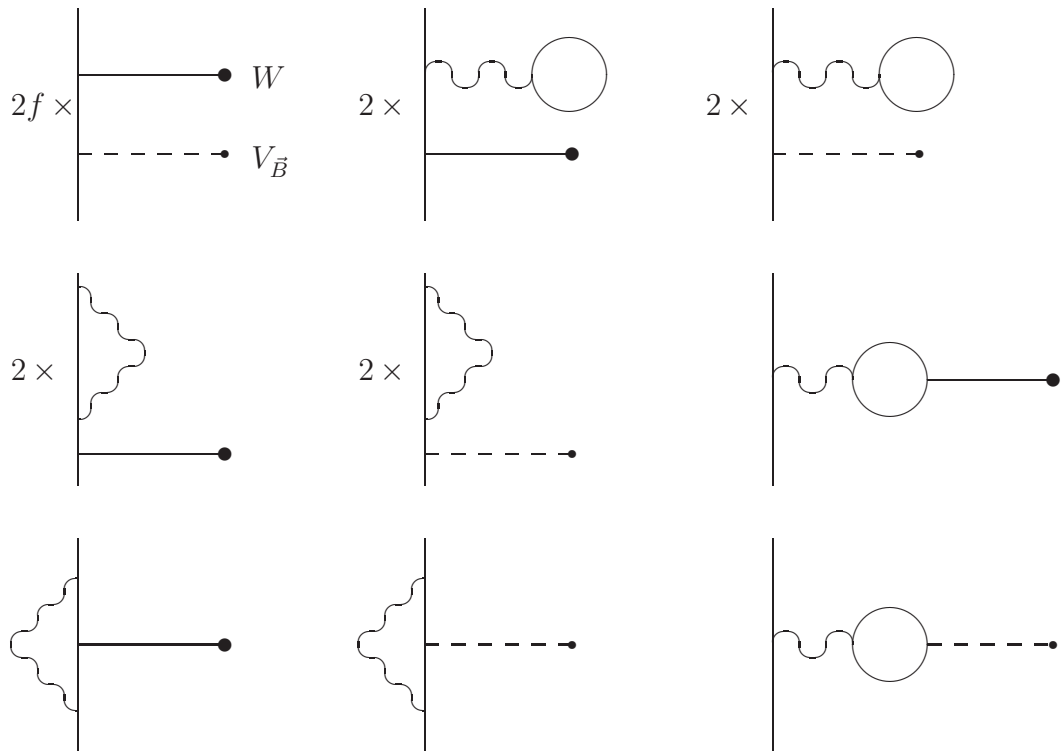
\begin{figure}
\begin{picture}(460,280)
\multiput(50,200)(130,0){3}{\line(0,1){80}}
\put(50,255){\line(1,0){55}}
\put(105,255){\circle*{5}}
\multiput(50,225)(10,0){6}{\line(1,0){5}}
\put(105,225){\circle*{3}}
\put(25,240){$2f\,\times$}
\put(115,250){$W$}
\put(115,220){$V_{\vecb}$}
\multiput(185,255)(20,0){2}{\oval(10,10)[t]}
\multiput(195,255)(20,0){2}{\oval(10,10)[b]}
\put(234,255){\circle{29}}
\put(180,225){\line(1,0){55}}
\put(235,225){\circle*{5}}
\put(155,240){$2\,\times$}
\multiput(315,255)(20,0){2}{\oval(10,10)[t]}
\multiput(325,255)(20,0){2}{\oval(10,10)[b]}
\put(364,255){\circle{29}}
\multiput(310,225)(10,0){6}{\line(1,0){5}}
\put(365,225){\circle*{3}}
\put(285,240){$2\,\times$}
\multiput(50,100)(130,0){3}{\line(0,1){80}}
\put(50,170){\oval(10,10)[tr]}
\put(50,130){\oval(10,10)[br]}
\put(60,170){\oval(10,10)[bl]}
\put(60,130){\oval(10,10)[tl]}
\put(60,160){\oval(10,10)[tr]}
\put(60,140){\oval(10,10)[br]}
\put(70,160){\oval(10,10)[bl]}
\put(70,140){\oval(10,10)[tl]}
\put(70,150){\oval(10,10)[tr]}
\put(70,150){\oval(10,10)[br]}
\put(50,112){\line(1,0){55}}
\put(105,112){\circle*{5}}
\put(25,140){$2\,\times$}
\put(180,170){\oval(10,10)[tr]}
\put(180,130){\oval(10,10)[br]}
\put(190,170){\oval(10,10)[bl]}
\put(190,130){\oval(10,10)[tl]}
\put(190,160){\oval(10,10)[tr]}
\put(190,140){\oval(10,10)[br]}
\put(200,160){\oval(10,10)[bl]}
\put(200,140){\oval(10,10)[tl]}
\put(200,150){\oval(10,10)[tr]}
\put(200,150){\oval(10,10)[br]}
\multiput(180,112)(10,0){6}{\line(1,0){5}}
\put(235,112){\circle*{3}}
\put(155,140){$2\,\times$}
\multiput(315,140)(20,0){2}{\oval(10,10)[t]}
\put(325,140){\oval(10,10)[b]}
\put(354,140){\circle{29}}
\put(368,140){\line(1,0){45}}
\put(415,140){\circle*{5}}
\multiput(50,0)(130,0){3}{\line(0,1){80}}
\put(50,60){\oval(10,10)[tl]}
\put(50,20){\oval(10,10)[bl]}
\put(40,60){\oval(10,10)[br]}
\put(40,20){\oval(10,10)[tr]}
\put(40,50){\oval(10,10)[tl]}
\put(40,30){\oval(10,10)[bl]}
\put(30,50){\oval(10,10)[br]}
\put(30,30){\oval(10,10)[tr]}
\put(30,40){\oval(10,10)[tl]}
\put(30,40){\oval(10,10)[bl]}
\put(50,40){\line(1,0){55}}
\put(105,40){\circle*{5}}
\put(180,60){\oval(10,10)[tl]}
\put(180,20){\oval(10,10)[bl]}
\put(170,60){\oval(10,10)[br]}
\put(170,20){\oval(10,10)[tr]}
\put(170,50){\oval(10,10)[tl]}
\put(170,30){\oval(10,10)[bl]}
\put(160,50){\oval(10,10)[br]}
\put(160,30){\oval(10,10)[tr]}
\put(160,40){\oval(10,10)[tl]}
\put(160,40){\oval(10,10)[bl]}
\multiput(180,40)(10,0){6}{\line(1,0){5}}
\put(235,40){\circle*{3}}
\multiput(315,40)(20,0){2}{\oval(10,10)[t]}
\put(325,40){\oval(10,10)[b]}
\put(354,40){\circle{29}}
\multiput(368,40)(10,0){5}{\line(1,0){5}}
\put(415,40){\circle*{3}}
\end{picture}
\caption{The second-order diagrams 
contributing to $S_2^{(\rm t)}(\az)$ and 
$U_2^{(\rm t)}(\az)$ being multiplied by the 
first-order diagrams presented 
in Fig. \ref{Diagrams31}.}\label{Diagrams32} 
\end{figure}
\end{document}